\documentclass{aa}
\usepackage{graphicx}
\usepackage{txfonts}
\begin{document}
   \title{3-Dimensional Kinematics in low foreground extinction
     windows of the Galactic Bulge}

   \subtitle{Radial Velocities for 6 bulge fields: \\
Procedures and Results \thanks{Based on observations
       collected at the European Organisation for Astronomical
       Research in the Southern Hemisphere, Chile.  European Southern
       Observatory (ESO) programme numbers 71.B-3048(A), 077.B-0600(A)
       and 079.B-0232(A).} 
       \thanks{Based on observations made with the NASA/ESA Hubble
         Space Telescope, obtained at the Space Telescope Science
         Institute, which is operated by the Association of
         Universities for Research in Astronomy, Inc., under NASA
         contract NAS 5-2655. These observations are associated with
         proposals GO-8250, GO-9436, GO-9816 and GO-11655.} 
}

   \author{M.Soto
          \inst{1,}
          \inst{3},
           K.Kuijken
          \inst{1}
          \and
          R.M. Rich\inst{2}%\fnmsep\thanks{Just to show the usage
%          of the elements in the author field}
          }

   \institute{Leiden Observatory, Leiden University
              PO Box 9513, 2300RA Leiden, The Netherlands 
              \email{ kuijen@strw.leidenuniv.nl}
         \and
             Department of Physics and Astronomy, UCLA, 
             Los Angeles, CA, 90095-1547
             \email{rmr@astro.ucla.edu}
%             \thanks{The university of heaven temporarily does not
%                     accept e-mails}
         \and 
            Departamento de F\'{\i}sica, Universidad de la Serena,
            Benavente 980, La Serena, Chile 
            \email{msoto@dfuls.cl}          
             }

   \date{Received January 15, 2011; accepted January 27, 2012}

% \abstract{}{}{}{}{} 
% 5 {} token are mandatory
 
  \abstract
   {} %leave it empty if necessary  
   {
     The detailed structure of the Galactic bulge still remains
    uncertain. The strong difficulties of obtaining 
    observations of stars in the 
    Galactic bulge have hindered the acquisition of a kinematic representation
    for the inner kpc of the Milky Way. The observation of the 3-d kinematics 
    in several low foreground extinction windows can solve this problem.
   }
   {
   We have developed a new technique, which combines precise stellar HST 
  positions and proper motions with integral field spectroscopy, in order 
  to obtain reliable 3-d stellar kinematics in crowded fields of the 
  Galactic center.  
    }
   { 
    In addition, we present results using the new techniques for six 
  fields in our project. 
  A significant vertex deviation has been found  
  in some of the fields in agreement with previous determinations.
  This result  confirms
  the presence of a stellar bar in the Galactic bulge.
   }
   {}


 \titlerunning{Radial Velocities for 6 bulge fields} 

   \maketitle
%
%________________________________________________________________

\section{Introduction}

The Milky Way bulge is the nearest example of a bulge/spheroidal population
that we can observe. Its proximity allows us to resolve stellar populations
and the associated kinematics, something which is not possible in external
galaxies. Even though many data have been gathered, a detailed 
unified  picture of the Galactic bulge including abundances, stellar 
populations and kinematics is far from being completed. 

 One of the main difficulties is the
location of the Sun inside the disk dust layer, which
limits observations to a few windows where the foreground dust extinction is
relatively low. In addition, populations in these windows are projected
on top of each other, complicating the analysis. 
  Disk and bulge components 
 overlap in the color-magnitude diagram specially near the turn-off (\cite{holtzman}),
 hampering a selection based on photometric criteria alone.   

\begin{figure*}[t]
\centering
\includegraphics[width=12.0cm]{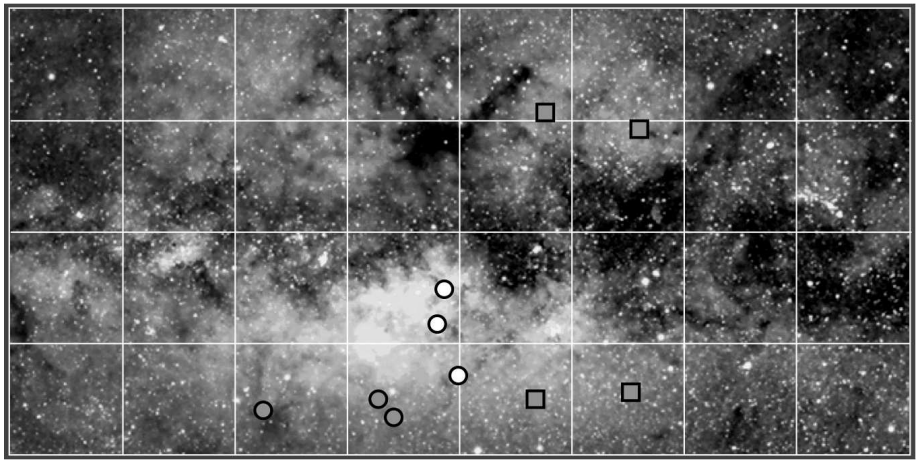}
\caption{ Fields in the Galactic Bulge observed for this project, 
 superimposed on an optical map (Mellinger 2008), 
 from longitude $+$20$^{\circ}$ to -20$^{\circ}$, and latitude -10$^{\circ}$ 
 to $+$10$^{\circ}$. White and grey
 circles correspond to fields for which proper motion and radial 
 velocity measurements have been completed. Data sets 
 for the four fields at negative longitudes (grey squares) have not 
 been completed so far.}
\label{fig:fields} 
\end{figure*}
 
 In spite of these limitations, important information has been gathered
 over the years. One of the pioneering studies of the kinematics of 
 the Galactic bulge was that of Spaenhauer et al. (1992), who measured proper
 motions for $\sim$400 stars from photographic plates obtained in 1950 and 1983. 
 This proper motion sample was 
 the basis for subsequent abundance and radial velocity studies 
 of the original proper motion sample (Terndrup et al. 1995, Sadler et al. 1996). 
 Zhao et al. (1994) combined the results of these studies with those 
 obtained previously by Rich (1988, 1990), and compiled a small subsample of 
 62 K Giants with 3-d kinematics and abundances. In spite of
 its small size the subsample showed a significant vertex deviation, 
 a signature of barlike kinematics. This result
 has recently been confirmed with a larger sample 
 of $\sim$300 stars (Soto, Rich \& Kuijken 2007). 
 de Vaucouleurs (1964) had originally 
 suggested that our Galactic bulge was actually barred,
 based on the similarity  of its spiral structure with other galaxies 
 with strong bars. Nevertheless, direct stellar signatures of the barlike 
 structure had not been found before. 
 Other studies have also produced important kinematic catalogs for the
 regions at the center of our galaxy. Sumi
 et al. (2004) used the information provided by the OGLE catalog to 
 produced $5 \times 10^6$ stellar proper motions in 49 bulge fields. 
 The latter results have been the basis of a study of proper motion
 trends in the Red Clump of the bulge region (Rattenbury 2007a; 2007b).  
 The Rattenbury et al. (2007a) selection, however, seems to have
 suffered from a significant disk contamination as pointed out by
 other authors (Vieira et al. 2007). Plaut's Window $(l, b = 0^{\circ}, −8^{\circ})$
 has also provided proper motions (Vieira et al. 2007), which
 have been obtained with plate observations spanning 21 years of 
 epoch difference. Bulge stars in these fields have been selected by 
 cross-referencing with the 2MASS catalog in order to obtain a clean
 sample of bulge giants. More recently, Clarkson et al. (2008)
 has produced  a new catalogue of $\sim 180,000$ proper motion using  
 two ACS epochs; from this sample $> 15,000$  bulge proper motions
 were  identified using a kinematic selection.

 In addition to models of the stellar distribution (e.g. Zhao 1996) 
 gas observations and hydrodynamical models 
 also have been used to study the Galactic bulge 
 (e.g. Englmaier \& Gerhard 1999).
 Many of these models rely on three dimensional deprojections of the 
 Galactic bulge  derived from the COBE DIRBE images (Dwek et al. 1995) whose results showed 
 asymmetries consistent with a  stellar bar in the Galactic center. 
 Even though all analyses agree on the rough orientation of the bar,  
 complete agreement about the
 values of the parameters which would define this bar, such as 
 rotational bar pattern speed or position angle has not been reached yet.
 For example, values for the angle between the bar's major axis and 
 our line of sight to the Galactic center have ranged from
 $\sim$10$^{\circ}$  (e.g. Picaud \& Robin 2004 and references therein)
 to 80$^{\circ}$  (Collinge et al. 2006). Furthermore, the presence of
 2 bar structures coexisting in the Galactic bulge has 
 been proposed by some authors  
 (Cabrera-Lavers et al. 2008 and references therein)
 ; the two more common
 bar angles ($\sim$20$^{\circ}$ and $\sim$40$^{\circ}$) belonging 
 to 2 distinct triaxial structures,  a boxy bulge/bar and a
 more extended long bar respectively.
 Nevertheless, this general picture with two bars rotating at different angles 
 is dynamically complex (Martinez-Valpuesta \& Gerhard 2011), and
 therefore is still under debate. 

 Similarly, the recent discovery of two coexistent 
 red clumps in the Galactic bulge on 2MASS and OGLE-III data (McWilliam \&
 Zoccali 2010; Nataf et al. 2010) and observed at high latitudes, has
   been interpreted as  the effect of an X-shaped structure
   in the bulge (Saito et al. 2011).
  Hence, all this evidence suggest a complex bulge
   structure with several components detected at different lines of sight.

 Understanding the bulge 
 kinematics requires understanding the gravitational potential that drives the 
 orbits (Kuijken 2004, henceforth K04). Once the kinematics are understood,
 they can be correlated with stellar population information 
 to build a picture of the galaxy evolution and bulge formation scenario.

 In order to improve our knowledge of the stellar kinematics in the bulge region 
 we have embarked on a project to obtain three-dimensional velocities 
 for a large sample of bulge stars, by combining HST proper motions measurements
 with VIMOS spectroscopy. 

 In this paper we present integral-field (IFU) spectroscopic 
 measurements for six bulge fields that have HST proper motion
 measurements: three fields on the minor axis 
 (Kuijken \& Rich 2002, henceforth KR02; KR04) as well
 as three fields at positive longitudes.   
 We have combined the IFU data cubes with photometric information in a new 
 procedure designed to work in crowded fields; the technique combines the 
 precise HST photometry and IFU spectroscopy to optimize the spectral 
 extraction. 

  Stellar kinematics involves the measuring the phase-space
 distribution function. This phase space generally 
 has three degrees of freedom. By providing 4-6 coordinates per star (the
 two proper motions, two sky coordinates, a distance determination by means
 of a main sequence photometric parallax, and a radial velocity for a subsample
 of bright stars) we can overconstrain the phase-space distribution 
 which will allow us a reliable determination of the orbit structure.

 The outline of this paper is as follows. In section 2 we will briefly 
 explain the project of which the work presented in this paper is a part, section
 3 is an account of the observations and the methods
 involved in each case. Section 4 contains the results of our analysis. 
 Finally section 5 is the summary and conclusions for this work.
  
\begin{table*}
\begin{center}
\caption{Radial Velocity and Proper-Motion Fields\label{tab:propfi}. 
}
%\footnotesize
\begin{tabular}{ l  l  c c c}
\hline
\hline
Field               &   PM Epoch     & PM Instrument &       (l,b)   &   $\alpha,\delta$ (J2000.0) \\
\hline				    		                   
Sgr-I               &   1994 Aug     & WFPC2      &   (1.26, -2.65)  &  17 59 00, -29 12 14 \\
                    &   2000 Aug     & WFPC2      &                  &  \\
Baade's Window      &   1994 Aug     & WFPC2      &   (1.13, -3.76)  &  18 03 10, -29 51 45 \\
                    &   1995 Sep     & WFPC2      &                  &  \\
                    &   2000 Aug     & WFPC2      &                  &  \\
NEAR NGC 6558            &   1997 Sep     & WFPC2      &   (0.28, -6.17)  &  18 10 18, -31 45 49 \\
                    &   2002 Aug     & WFPC2      &                  &  \\
Field 4-7           &   1995 Jul     & WFPC2      &   (3.58, -7.17)  &  18 22 16, -29 19 22\\
                    &   2004 Jul     & ACS/WFC    &                  &  \\
Field 3-8           &   1996 May     & WFPC2      &   (2.91, -7.96)  &  18 24 09, -30 16 12\\
                    &   2004 Jul     & ACS/WFC    &                  &  \\
Field 10-8          &   1995 Sep     & WFPC2      &   (9.86, -7.60)  &  18 36 35, -23 57 01\\
                    &   2004 Jul     & ACS/WFC    &                  &  \\
\hline
\end{tabular}
\end{center}
\normalsize
\end{table*}

\section{Project}

 The HST data archive contains a treasure in WFPC2 images taken during the 
 1990s. 
 This wealth of images can be used to find suitable first epoch fields for 
 proper motion work; we have chosen ten for this project, with low
 foreground extinction, sufficiently deep exposures, and spread 
 in $l$ and $b$. 
 Hence, the HST archive has provided us with first epoch observations in six fields 
 at $l\sim 0$, and $l>0$; in addition, we have established four fields at $l<0$ in order
 to target both ends/sides of the bar/bulge.
 The goals for each field are the acquisition of color magnitude diagrams, accurate 
 astrometry, and radial velocities for as many stars as possible.
 
  Figure \ref{fig:fields} shows all the fields for this project.
%  superimposed on an optical map (Mellinger 2008). 
  HST archive images were primarily used to set first epoch proper motion
 exposures in several low extinction bulge regions, close to the Galactic minor 
 axis and at positive longitudes. These initial fields were 
 complemented more recently with observations in 
 four more fields at negative longitudes. Thus, this project strategically 
 spans a wide range of bulge locations, sampling a significant stellar
  population at the center and both sides of the bulge/bar. 
 Consequently, the proper motion results published in KR02 and K04, 
 represent the first important
  piece of kinematic information on this project, which we continue here. 
 The complete HST programme described before, which points to proper motions, 
 photometry and parallax distances has been more recently combined 
 with a spectroscopic 
 VLT programme in the same fields, this spectroscopic information, and the techniques involved
 are the subject of this paper, where Table \ref{tab:propfi} shows the coordinates 
 of each field.

%_________________________________________________________________

\section{Observations and procedures}

\subsection{Proper motions}  
  First epoch photometric observations with WFPC2 for all
  the fields used in this paper were obtained from the Hubble Space 
  Telescope data archive. 
 In the case of the three fields close to the Galactic minor axis 
 (near $l=$0$^{\circ}$) second epoch observations over a time baseline
 of 6 years have resulted in accuracies better than 1 mas/yr, 
 which corresponds to
 errors below 30 km/sec at the distance of the bulge, significantly 
 smaller than the velocity dispersion of the bulge of 100 km/sec.
 Even longer time baselines for the fields at positive longitudes were used
 (8-9 years) as Table~\ref{tab:propfi} shows.  
 First and second epochs were taken with WFPC2 for fields close to the 
 Galactic minor axis, conversely fields at positive longitudes 
 used a combination of WFPC2 and ACS for first and second epoch
 respectively. The latter fields thus had to include small differences
 in the procedure to take into account the instrument change (e.g. the 
 shearing of ACS images with respect to WFPC2).

\begin{figure}
\begin{center}
\includegraphics[width=9cm]{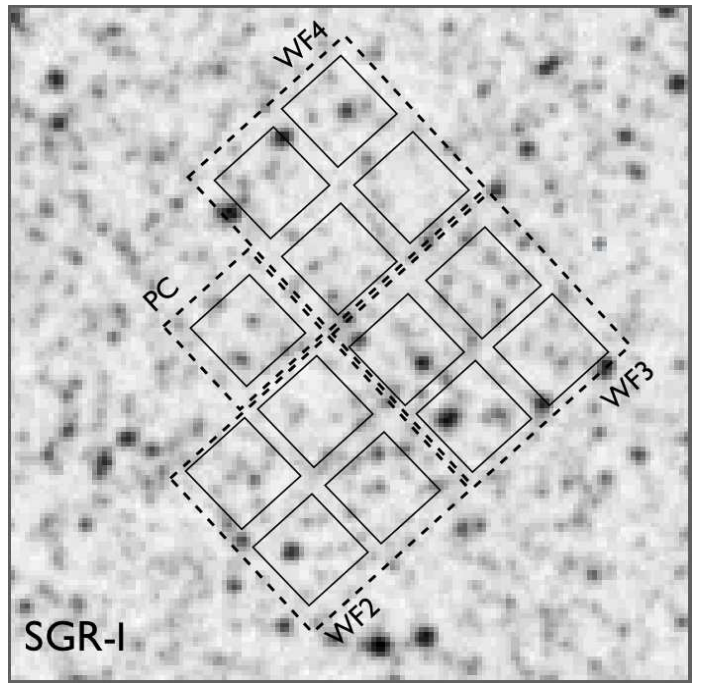} 
\caption{ Finding chart for one of our fields, \emph{Sagittarius-I}, using an image from
 2MASS. Each small square (solid line) corresponds to each one of the VIMOS IFU fields. 
 Dashed squares correspond to PC, WF2, WF3 and WF4 HST fields superimposed on the same 
 image.
 \label{fig:mychart}}
\end{center}
\end{figure}

 Proper motions were measured using a modification of the Anderson \& King 
 (2000) procedure, which consists of a combination of PSF reconstruction
 and PSF core fitting (KR02). 
 A more detailed account 
 about the proper motion measurements can be found in KR02 and it will not be 
 repeated here. 

\subsection{Radial Velocities}

 The procedure to obtain the spectrum of each star in these crowded
 fields consists of two main steps, the extraction of the spectra for 
 each fiber/pixel in the IFU field, and the extraction of the star
 spectra from the IFU data cube. During the second step we will 
 combine the spectroscopy with the information yielded by HST imaging.
 
 The VLT VIMOS Integral Field Unit (IFU) has a 27''$\times$27'' 
 field of view in high resolution (R $\sim $2050) which
 allows spectra to be taken on a 40 $\times$ 40 grid, of spacing 0.67''.
 This permits spectroscopy of a large number of bulge stars in a single
 exposure, at a resolution of $0.54\ \AA/pixel$ from $4150$ to $6200\  \AA$. 
 We used this instrument to target our HST fields, which can each be
 covered by 13 VIMOS pointings (4 per WF chip and 1 per PC, as Figure \ref{fig:mychart} 
 illustrates for Sagittarius-I). 
 Each IFU pointing was exposed for  2 $\times$ 1000 sec , which has
 allowed us to resolve approximately 80 stars per IFU field.
 The spectra 
 yield 30 $km \ s^{-1}$ radial velocity precision, which is well-matched
 to the transverse velocity accuracy from our proper motions
 (better than 1 $mas/yr$, equivalent to $\sim$30 $km \ s^{-1}$ at 8 kpc
 distance), and sufficient to resolve the velocity dispersion in the central 
 parts of the Galaxy, which is about 100 $km \ s^{-1}$.

 In addition to the regular science images containing the information
 about our six HST fields, we also made observations of 
 highly extincted bulge fields for use as "sky'' exposures.
 Standard stars
 were observed as well, for use as templates in the cross-correlation 
 process for the determination of the velocities.
 The overall observation time for all the spectral observations
 was 17, 50 and 45 hours for our three observing runs respectively.  
 Table~\ref{tab:veldate} summarizes the VIMOS IFU observations for 
 the six fields presented in this paper. In Table~\ref{tab:veldate}  
 the numbers under every
 ``run'' column correspond to the number of IFU fields observed in that run.
 All data was taken in service mode with seeing conditions constrained 
 at a maximum of 0.8''. 

\begin{table*}
\begin{center}
\caption{Summary of Radial Velocity Observations \label{tab:veldate}. 
}
\begin{tabular}{ l  c  c  c  c  c}
\hline
\hline
Field       & 1$^{st}$ run  & 2$^{nd}$ run & 3$^{rd}$ run  & Total IFU Fields  & Stars with Rad. Vel.\\
            &  (2003) & (2006) & (2007) & & \\
\hline
Sgr-I               &    5  &   6  &  5  & 16 & 962 \\
Baade's Window      &    5  &   5  &  4  & 14 & 965 \\
NEAR NGC 6558            &    5  &   5  &  4  & 14 & 766 \\
Field 4-7           &    0  &   8  &  3  & 12 & 664 \\
Field 3-8           &    0  &  10  &  3  & 13 & 466 \\
Field 10-8          &    0  &   9  &  4  & 13 & 756 \\
\hline
\end{tabular}
\end{center}
\end{table*}

\begin{figure}
\begin{center}
\includegraphics[width=8.5cm]{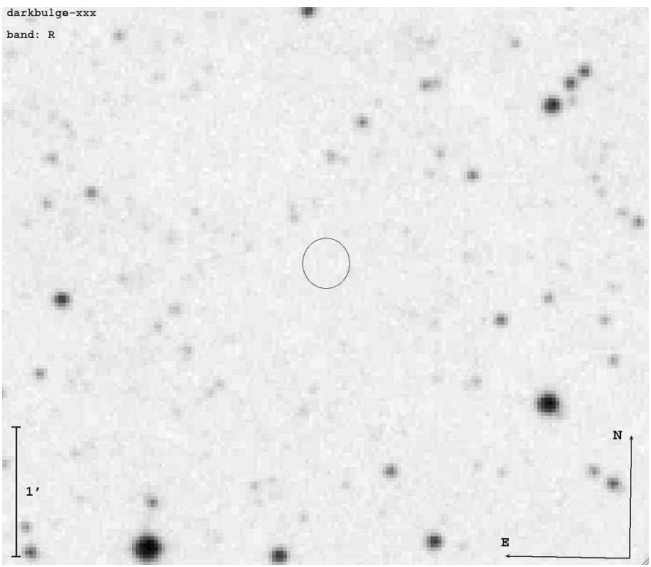}
\caption{Finding chart of one of our darkbulge fields (circle in the middle) 
 over a 2MASS image. 
 Darkbulge fields have been used during the sky subtraction in 
 VIMOS IFU fields. 
\label{fig:darkbulge}}
\end{center}
\end{figure}

\subsubsection{Data cube organization and Radial velocity measurements}
 
 VIMOS IFU raw data are complex to reduce and calibrate.
 Fiber spectra extraction was carried out using the ESO pipeline 
 for VIMOS IFU data. Programs GASGANO\footnote{available at http://www.eso.org/sci/software/gasgano/} and ESOREX\footnote{available at http://www.eso.org/sci/software/cpl/esorex.html} were used to manage 
 the VIMOS IFU recipes\footnote{available at http://www.eso.org/projects/dfs/dfs-shared/web/vimos/vimos-pipe-recipes.html} (Details about methods and procedures of the
 recipes can be found in VIMOS pipeline User's Guide 
 and Gasgano User's Manual). The recipes 
 used during our processing were \emph{vmifucalib} and \emph{vmifuscience}.   
 
 The final product of the VIMOS IFU recipes are the spectra extracted and 
wavelength calibrated in one image that includes all the spectra for each 
quadrant in the IFU field.  

  An important problem to be considered in spectroscopic reduction 
 is related to the sky 
 subtraction, which has not been implemented by the VIMOS pipeline 
 (VIMOS Pipeline User's Guide 7.23.11). We have tried 
 two approaches. The first is basically 
 the same recommended by the VIMOS Pipeline User's Guide. We took the 20 
 spectra with lowest signal per quadrant (which means 5\% of the total) 
 and averaged them (taking care to reject the dead fibers). 
 The combined spectrum 
 was considered as sky and subtracted from the rest of the fibers. This way
 of proceeding is not ideally suited to crowded fields and could
 change the results by subtracting a flux level too high (or too low) from the reduced 
 spectra. 
 The second method involves exposures of nearby highly extincted 'dark bulge'
 fields, whose spectra, appropriately scaled, mimic the sky contribution to 
 the stellar fields as Figure \ref{fig:darkbulge} shows.

 Both processes were extensively tested to check their influence on our radial 
 velocity results; we found no significant differences for both procedures,
 typically below 3 $km \ s^{-1}$ in the final velocity measurements per fiber. 
 Given the reliability of our extraction we have preferred to use
 the sky extraction by dark fields in our fields.

\begin{figure}
\centering
\includegraphics[width=9.0cm]{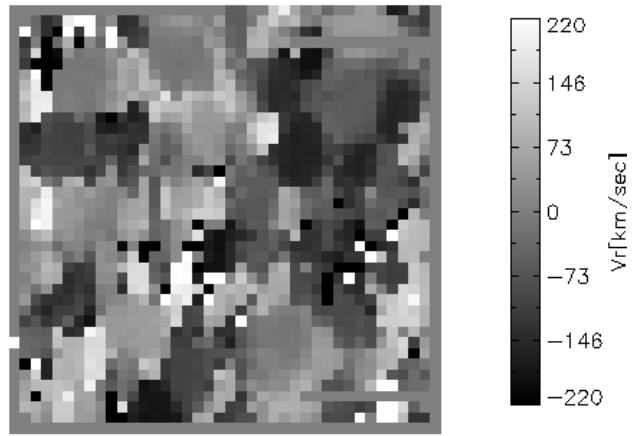}
\caption{
 Velocity field for one of our IFU observations in Baade's window.
 The velocity for each pixel/fiber has been calculated using   
 cross-correlation, where each fiber corresponds to 0.66'' \label{fig:velres}.
 The VIMOS-IFU instrument allows clearly to distinguish between adjacent 
 stars with different kinematics.
}
\end{figure}

 Once the spectra were reduced we assembled them into spectral data cubes.
 In addition 
 to the regular calibrations, we produced for each IFU field, a response map
 to check the normal behavior of the fibers through the field. Dead 
 fibers or lost traces are easily highlighted in this way.

 The last step is the radial velocity measurement per fiber in each IFU 
 field. The measurement of radial velocities was made in all cases
 with a cross correlation in the IRAF\footnote{"IRAF is distributed by the National Optical
 Astronomy Observatories, which are operated by the Association of
 Universities for Research in Astronomy, Inc., under cooperative agreement
 with the National Science Foundation."} task \emph{fxcor} (\cite{tonry79}) 
 against a template standard star (HD157457). Our template star  
 has a Spectral type  G8III, and therefore we expect a negligible bias
 favoring similar spectral types. 
 Before the cross correlation some zones 
 of the spectra were masked; for instance the atmospheric emission line 
 due to OI at 5577.5 \AA; 
 or the interstellar absorption NaD lines at 5889 \AA.
 The latter lines are particularly strong in K2-3 III Giants, and can
 easily bias the correlation with our template to calculate velocities or
  a possible spectral type classification. 
 An example of the velocity field calculated with the procedure here 
 described for one of our IFU fields in Baade's Window 
 appears in Figure~\ref{fig:velres}, and
 clearly shows distinct colored zones which correspond to different
 stars at different velocities. Extracting this information optimally is 
 the subject of the next section.

\subsubsection{Deconvolution}
 
\begin{figure*}%[!t]
\centering
\includegraphics[width=15cm]{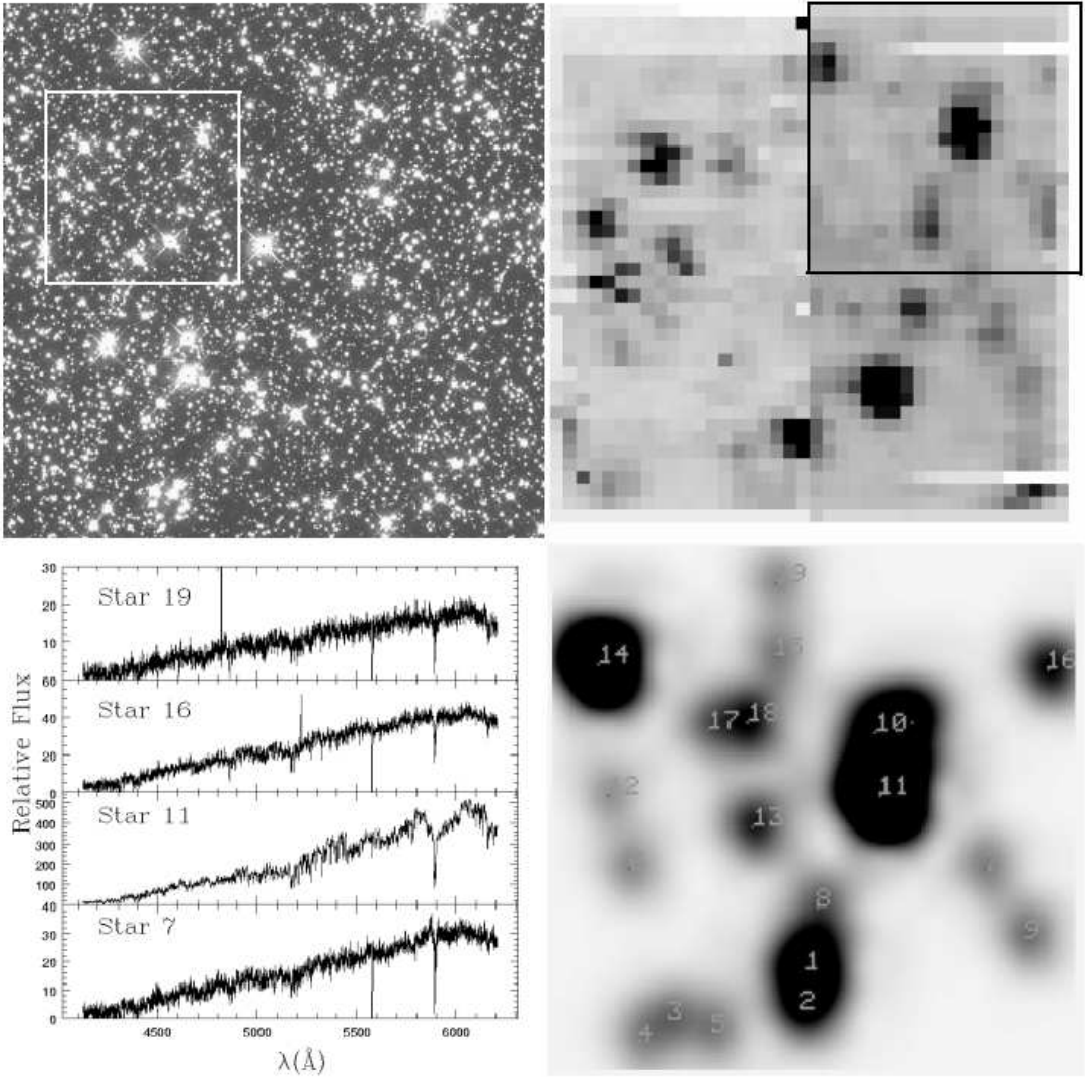}
\caption{Steps during the process to build a star spectrum from the 
 spectral cube. The top left figure corresponds to one of our observations
 with HST WF2 in Baade's window, the white square corresponds to the area
 covered by one of the VIMOS IFU images (top right). In the IFU field the first
 quadrant is enclosed, its respective convolved image produced during
 the deconvolution process to check the detection of stars in the first 
 quadrant appear at the bottom right. Finally some examples of the spectra 
 extracted by this process are shown at the bottom left. 
 \label{fig:deconv}  }   
\end{figure*}

 As shown in Figure \ref{fig:deconv}, the bulge fields are very crowded, 
 and therefore, a reliable technique to correct IFU spectra cubes for 
 blending is absolutely necessary.
 Fortunately, our HST images allow us to know precisely where
 the stars are. This information makes feasible the 
 optimal extraction of star spectra from the IFU cubes.

 The first step to carry out the deconvolution process is a coordinate 
 transformation from HST to IFU coordinates, 
 which is performed using the standard IRAF  
 tasks \emph{daofind} and \emph{geomap}.
 Since not all the stars observed in the HST image
 are detected in the
 IFU field, a threshold  magnitude for the HST list stars must be given.
 In the case of the fields Sgr-I, and BW, for which we have F555W observations,
 this limiting magnitude was set at V(F555W)$_{lim}$=21 mag. Similarly,
 for fields NEAR NGC6558 (which is actually located close to NGC 6558), 
 Field 4-7, Field 3-8, and Field 10-8
 the limiting magnitude is  V(F606W)$_{lim}$=20-20.5 mag, depending
 of the particular crowding in the IFU field. Using this limiting magnitude 
 we avoid the deconvolution of the complete list of stars detected in the 
 HST image, where naturally a large fraction of them are beyond 
 the detection limit of the VIMOS IFU observations.

 Armed with a magnitude-limited list of HST proper motions, positions, and magnitudes,
 we have produced a procedure which accounts for defects in the HST list to 
 separate the fluxes of as many stars as possible from the VIMOS IFU spectral cube.
 Thus, the HST list, once cleaned from spurious stars due to failures in the DAOFIND 
 detection procedure at 20 $\sigma$, is used to perform the deconvolution of the 
 stellar spectra
 in the IFU cube. The deconvolution requires a precise IFU PSF and the HST positions 
 in the IFU field. Simultaneously, a convolved image using the HST magnitudes
 and positions and the IFU PSF is created during the deconvolution process. 
 This convolved image yields an estimation of
 blending for each star which is used to select stars with a limited amount of blending
 for the final list of stars with radial velocities after the deconvolution. 
 Hence, once stars have been deconvolved from the IFU-cube from a HST cleaned list,  
 we can measure our radial velocities.

 This deconvolution procedure can be described in more detail as follows: 
With the final list of cleaned HST stars lying in the respective
IFU field and the PSF of the latter we can estimate the contribution 
of a star \emph{s} to each pixel \emph{i}, which defines the model,

\begin{equation} 
P_i = \sum_{s} F_s C_{si}, 
\end{equation}

 where $P_{i}$ is the flux in each pixel, $F_s$ is the flux in the star $s$
 and $C_{si}$ corresponds to the contribution of that star $s$ to the pixel
 $i$ obtained using the IFU PSF.

\begin{eqnarray}   
\chi^2 & = & \sum_i (P_i(observed) - P_i(model))^2 \\
       & = & \sum_i (P_i(observed) - \sum_s F_s C_{si})^2,
\end{eqnarray}

which can be solved by requiring
\begin{equation}
\frac{\partial \chi^2}{\partial F_s}=0 \ \ \ \ \ \ \ \ \ \ \ \ \forall s. 
\end{equation}
This results in the matrix showed in eq. 5
\begin{equation} 
\sum_{s'} F_{s'} \left( \sum_i C_{si} C_{s'i} \right)= \sum_i P_i C_{si}, \label{deconv5}
\end{equation}
 which defines a square system. The solution of this matrix for each slice 
of the IFU data corresponds to the flux solution to that wavelength for each 
 star; repeating the procedure in every slice of the cube we are able to 
 reconstruct the spectrum of each star. 

 This technique, simple in theory, might yield a singular matrix in some cases,
 when stars are too close to each other in the HST image, and therefore
 can not be resolved in the IFU image. Furthermore, saturation and bleeding 
 would augment the probability of obtaining a singular matrix in eq. \ref{deconv5},
 as these effects produce multiple DAOFIND detections around saturated spots.  
 In order to avoid a singular matrix during the deconvolution procedure, we have
 implemented several solutions, which we describe below:
  (1) Due to the differences in the pixel scale and resolution between the HST WFPC2
 and the VIMOS IFU image 
 (pixel scale is 0.05'' for PC, and 0.1'' for WF, while each pixel/fiber is 0.66'' 
 in VIMOS-IFU)  it is expected that in many cases more than one HST 
 star will fall in one single IFU pixel. Even in those cases, the fluxes of
 two stars in the same IFU pixel can be separated as long as they can be 
 resolved as single stars using the IFU PSF and the HST positions. The IFU PSF 
 has a typical FWHM in our observations of 1.8 pixels. 
 Thus, we have found realistic to set a minimum blending radius
 for the stars in the HST list, 
 stars closer than 0.01 IFU pixel-scale have been considered as one during the 
 deconvolution, where fluxes have been added in those cases. 
 (2) Similarly, false
 detections due to bleeding and saturation in the HST list are discarded by
 comparing with the positions of stars in the IFU field. We cross matched both lists, 
 identifying the HST star position which is closer to the position detected 
 in the IFU field, and discarding the rest of the detections in the HST list inside 
 an avoidance radius. 
 Typically an avoidance radius of 0.5$\times$FWHM of the IFU PSF has been used.
  (3) Nevertheless, in spite of the two procedures just described, false DAOFIND 
 detections in the HST list have eventually appeared during the deconvolution.
 These false stars in the HST list were particularly common in long exposures, 
 where saturation and bleeding in bright stars left saturation many pixels 
 away of the central position of the star. 
 In order to solve this we have devised a simple local procedure which
 iteratively compares the convolved image generated by the HST position, 
 magnitudes, and the IFU PSF with the real IFU field image; stars in the HST list
 with fluxes 2 $\times \ background $
 of a ratio image (real IFU image divided by a simulated IFU convolved image) are 
 rejected and marked as false stars to generate a new convolved
 image and a new loop. This iterative process rapidly converges and effectively 
 cleans of false stars due to saturated pixels the HST list of stars. 
 (4)Finally, once the deconvolution is performed, a final selection is carried out, 
 as we mentioned before. This last selection is intended to avoid stars 
 blended and/or mixing different populations which due to our radial velocity 
 measurement technique, would deliver an average velocity as a result.
  Hence, we measure as many velocities as possible in single stars. 
 Only stars in which the central 
 positions had at least 70 \%  of the total flux of the pixel were finally 
 selected. Typical $S/N$ for stars with radial velocity errors
 below 50 $km\ s^{-1}$ was $\sim 2.4$ between $4500\AA$ and 
 $6000\AA$.
An example of this process and its results is illustrated 
 in Figure~\ref{fig:deconv}.

\begin{figure}
\centering
\includegraphics[width=8.5cm]{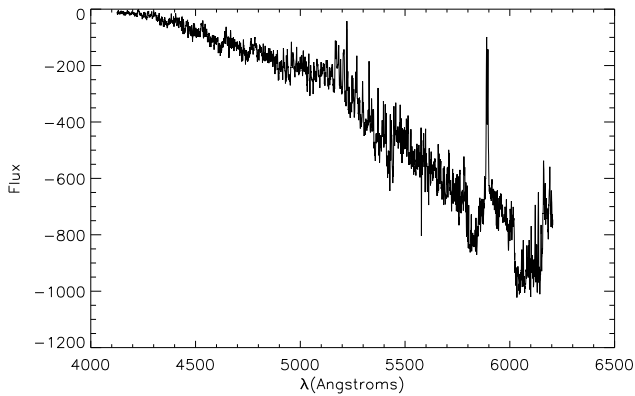}
\caption{ Example of the result of using a wrong PSF during the deconvolution.
 This is the deconvolved spectrum for the HST position in the star 11 in Figure 
 \ref{fig:deconv} obtained using a modified PSF. The flux overestimation
  in a neighbor star by the PSF produces, when the system is solved, 
 a negative solution for this star. A correlation can not be established 
 with these distorted spectra in the radial velocity measurements.
\label{fig:invspec}} 
\end{figure}

  A limitation of our technique is related to the PSF construction. 
 The IFU field is small as we 
 have already mentioned, which combined with the area covered for each
 fiber (0.66''), allows only a few detections in each IFU exposure
 ($\sim$40-80 typically, in normal conditions, with the four quadrants working).   
   This small number of detections often produces a heavily
   undersampled PSF, which is critical in the deconvolution, where this PSF is
 used as a model to estimate the flux of every star.
 
\begin{figure}
\centering
\includegraphics[width=9cm]{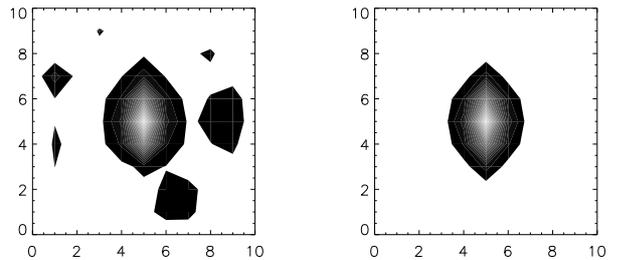}
\caption{Example of procedure to ``refit'' a IFU PSF. The undersampling of the 
 IFU PSF is fixed by a new fitting procedure which 
 gives more weight to central pixels in order to avoid background noise
 and contamination by neighbor stars. The new PSF is used during 
 the deconvolution process. 
 \label{fig:psfrefit} } 
\end{figure}

 During the development of our procedure each PSF was carefully obtained.
 The result of using a wrong PSF produces a flux overestimation of some stars.
 The result of this overestimation is negative fluxes in stars 
 in the neighborhood of some bright stars which when projected to 
 all the wavelengths in the spectral cube corresponds exactly to the spectrum
 of the neighbor bright star inverted as it is shown in Figure~\ref{fig:invspec}.
 A similar result is obtained when many HST positions are located in the 
 proximity of a very bright star, which is a typical failure of  
 photometric detection in bright saturated stars. In all these cases it is not 
 possible to obtain a reliable radial velocity measurement for these inverted spectra. 
 Limited solutions implemented in our code, and already described,
 relate and compare the stars detected in the HST and VIMOS IFU field in order
 to solve the crowding. The undersampling of the IFU PSF on the other hand
 requires additional information. 
 In order to improve this undersampled IFU PSF we performed, 
 a Gaussian ``\emph{refitting}'' of the IFU PSF generated by IRAF tasks, 
 where several PSF models were tested. 
 The PSF fit gives more weight to central pixels, where differences in 
 background flux between original and refitted PSF were typically of the 
 order of 3\% and did not show significant consequences in the final 
 velocity results.     
 An example of the latter procedure is shown in Figure~\ref{fig:psfrefit}, 
 and a summary of all radial velocity measurements for our the six fields 
 is indicated in table ~\ref{tab:veldate}.  

 Saturation not only affected our deconvolution procedure by 
 adding false stars in the HST DAOFIND, also it affected 
 the estimation of the amount of blending, and therefore the final 
 selection of stars in the IFU spectral cube.
 The problem of saturation affecting the values of the magnitudes, 
 was specially significant in NEAR NGC 6558
 and the three off-axis fields,
 which count with first epoch HST observations originally intended
 for the study of faint stars, and therefore with long exposure times.
 The latter problem was partially solved with the calculation 
 of \emph{m814} aperture photometry 
 magnitudes from short exposures (50 sec.)  
 of the ACS second epoch images. This was possible only for the 
 three off-axis fields Field 4-7, Field 3-8, and
 Field 10-8.
 Consistency of the velocity results has been 
 checked using both magnitudes, showing no differences and therefore 
 allowing us to avoid many misidentifications, after the deconvolution
 in these fields.

\subsubsection{An alternative deconvolution}
\begin{figure}
\centering
\includegraphics[width=8cm]{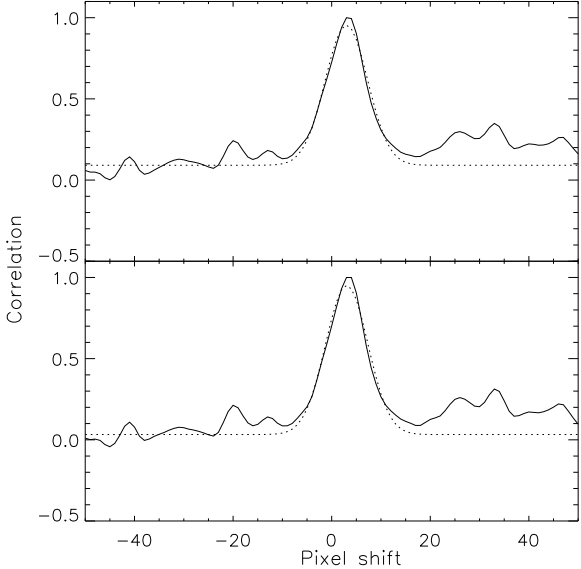}
\caption{Cross-correlation function for one of the stars in Baade's
  Window field. 
(\emph{Top}) Obtained from the velocity measurement of the deconvolved spectrum. 
(\emph{Bottom}) Obtained directly from the deconvolution of the CCF data cube.
 In each case a Gaussian fit has been performed (dotted
  line). The small differences between both CCF in this example produce output
  velocities with a difference of $2\ km\ s^{-1}$.
  \label{fig:ccf}} 
\end{figure}

 Since deconvolution and cross-correlation are both linear 
 operators, they can be carried out in any order. 
 Thus, an equivalent procedure to deconvolve the cross-correlation 
 function (CCF) obtained from the velocity measurements of all fibers in 
 the quadrant has been implemented. The procedure separates the contributions
 of the flux of each star in each pixel in the CCF data cube in the same way 
 that spectra for all the stars are deconvolved from the spectra data cube. 
 Consistency between the velocity results using both methods has been
 tested in the VIMOS field Baade's Window WF2c. We have
   measured an average modulus difference for this field of $7 \pm 8\ km\ s^{-1}$ which 
  we attribute mostly to small differences in the fitting of the CCF.
 Figure~\ref{fig:ccf} shows an example of the CCF from the same star in both cases; the 
 maximum in the deconvolved CCF
 is found at the same pixel, where each pixel corresponds to 31.78 km/sec.
 Thus, the deconvolution needs to be run on just a few pixels around the velocity
 zero channel (\emph{pixel shift}$= 0$) if the expected
 velocities are not too high (e.g. $|Vr| \leq 300\ km\ s^{-1}$).
 At high spectral resolution the CCF deconvolution could be an interesting 
 way to deblend data cubes spectrally.
 In spite of the advantages mentioned before, the CCF
  deconvolution technique was not
 used in the rest of the article; our radial velocity results are
 derived from the deconvolution of the data cubes containing stellar spectra.

\subsection{Zeropoint Velocity Corrections}

 Our data combine three observing runs as Table~\ref{tab:veldate} shows. 
 In order to check the reliability and performance of our technique
 we repeated in each observing run one of our fields in Sagittarius-I. \emph{Sgr1-pc} 
 has a reasonable crowding, with a lack of very bright stars, and 
 represents a typical example of the performance of VIMOS IFU during our bulge 
 observations.

\begin{figure*}
\centering
\includegraphics[width=6cm]{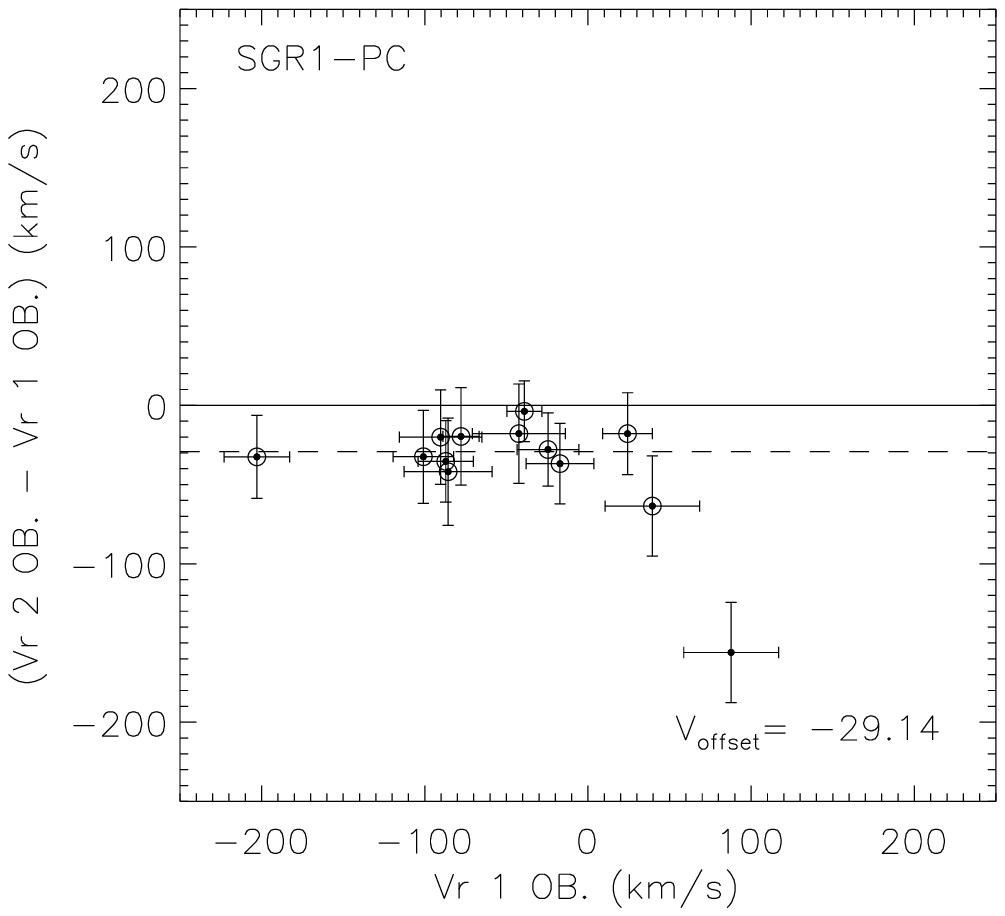}
\includegraphics[width=6cm]{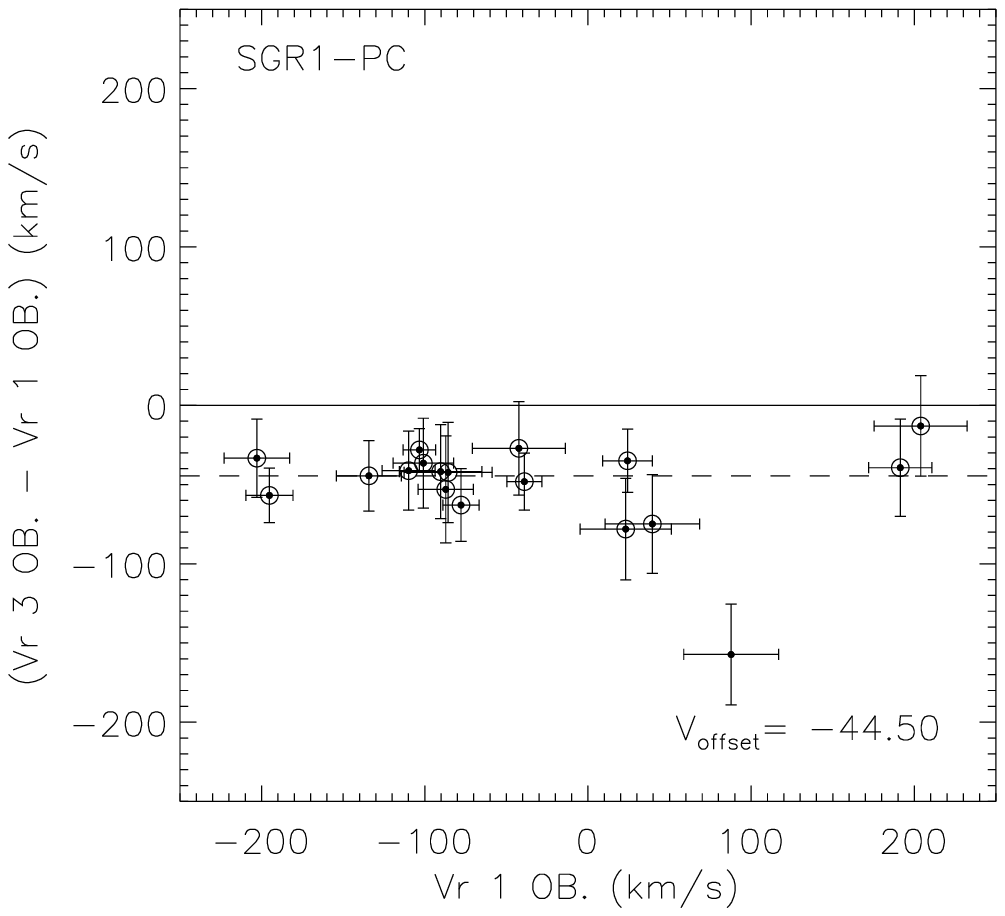}
\caption{ Velocity zeropoint determination for second and third run IFU fields. 
 IFU field \emph{Sgr1-pc} velocity results have been compared between the first 
 and second observing run (\emph{left}), and first and third (\emph{right}).
 The fit in each case is iterative, rejecting stars beyond 30 km s$^{-1}$. 
Those stars selected in the final iteration are enclosed
 by a circle. \label{fig:compvel}} 
\end{figure*}

 Each year's observations were analyzed with the same HST master list, and reduced
 independently. Figure~\ref{fig:compvel} shows a comparison of the radial velocities
 obtained. A significant offset is evident between the observing seasons. 
 Even though the origin of these deviations has not been 
 identified completely, we have corrected them in each case. 
 Accordingly, a zeropoint 
 offset has been added to the observations of the second and third observing runs, 
  29 $\pm$ 11 and 44 $\pm$ 8 (km s$^{-1}$)
 respectively, where errors have been estimated using 100,000
 bootstrap Monte Carlo realizations. These offsets were calculated in
 each case using an iterative clipping algorithm for all the stars in
 Sgr1-pc field 
 with radial velocities with measured errors below 30 (km s$^{-1}$).
 Furthermore, these offsets have been checked against the 5577.5 $\AA$ OI
 emission line which delivered consistent velocity offsets (28 $\pm$ 6
 and  38$\pm$6 for the second and third epoch
 respectively). On the contrary, a significant offset was not found for
 the first epoch. Further proof of the suitability of the applied
 offsets can be seen in \S 4. 

\section{Analysis}

\begin{figure*}
\centering
\includegraphics[width=6.5cm]{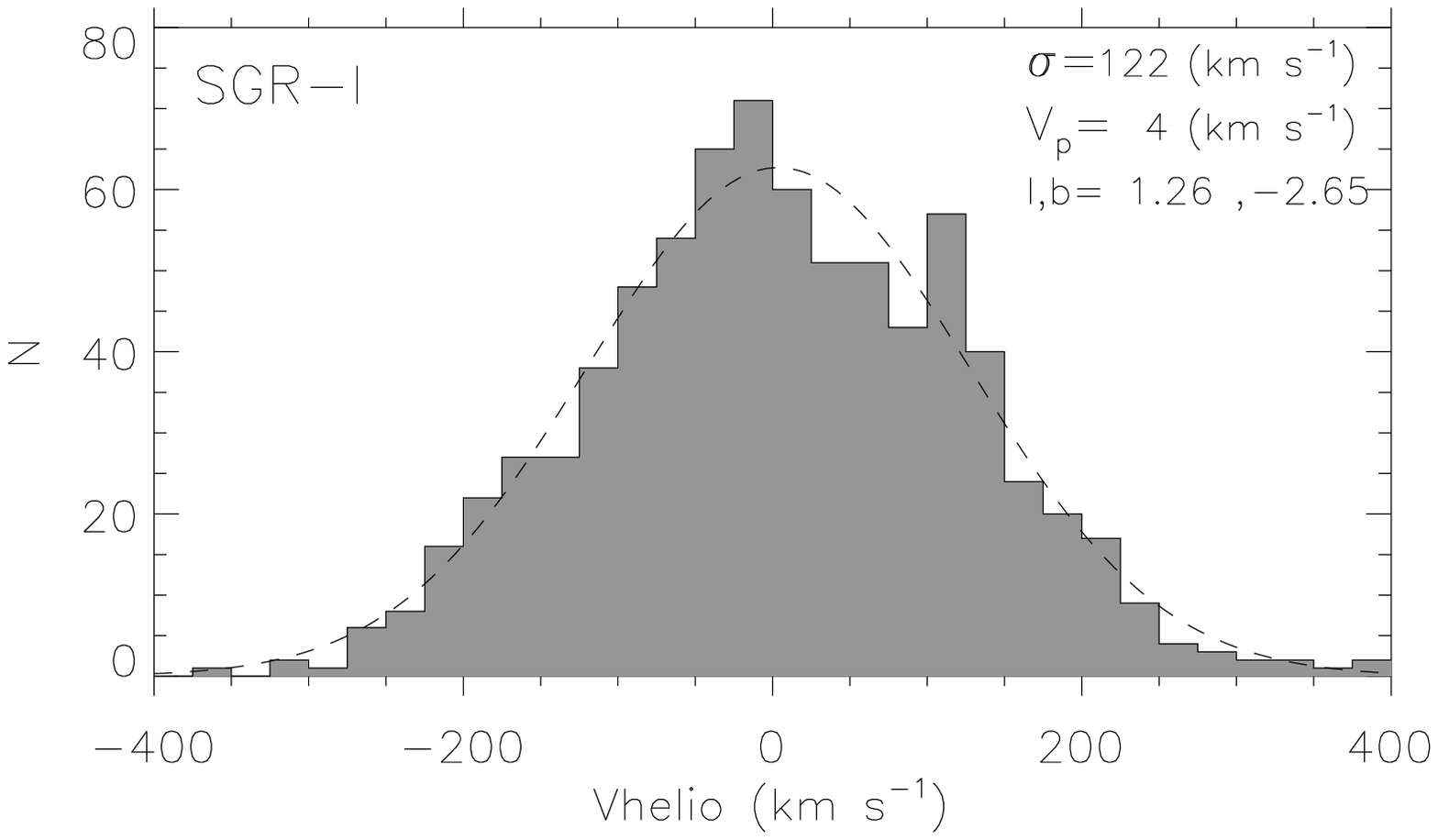}
\includegraphics[width=6.5cm]{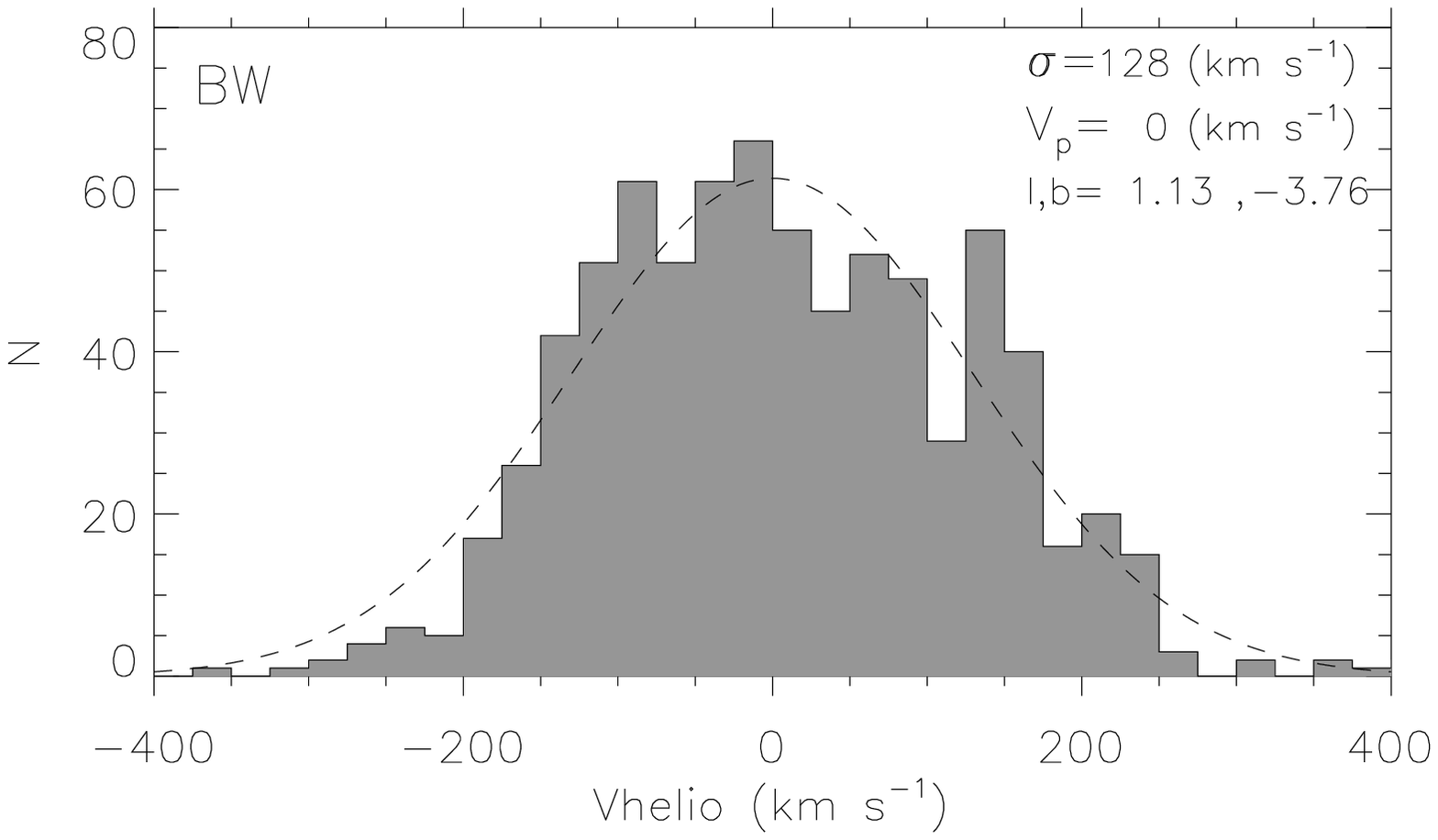}\\
\includegraphics[width=6.5cm]{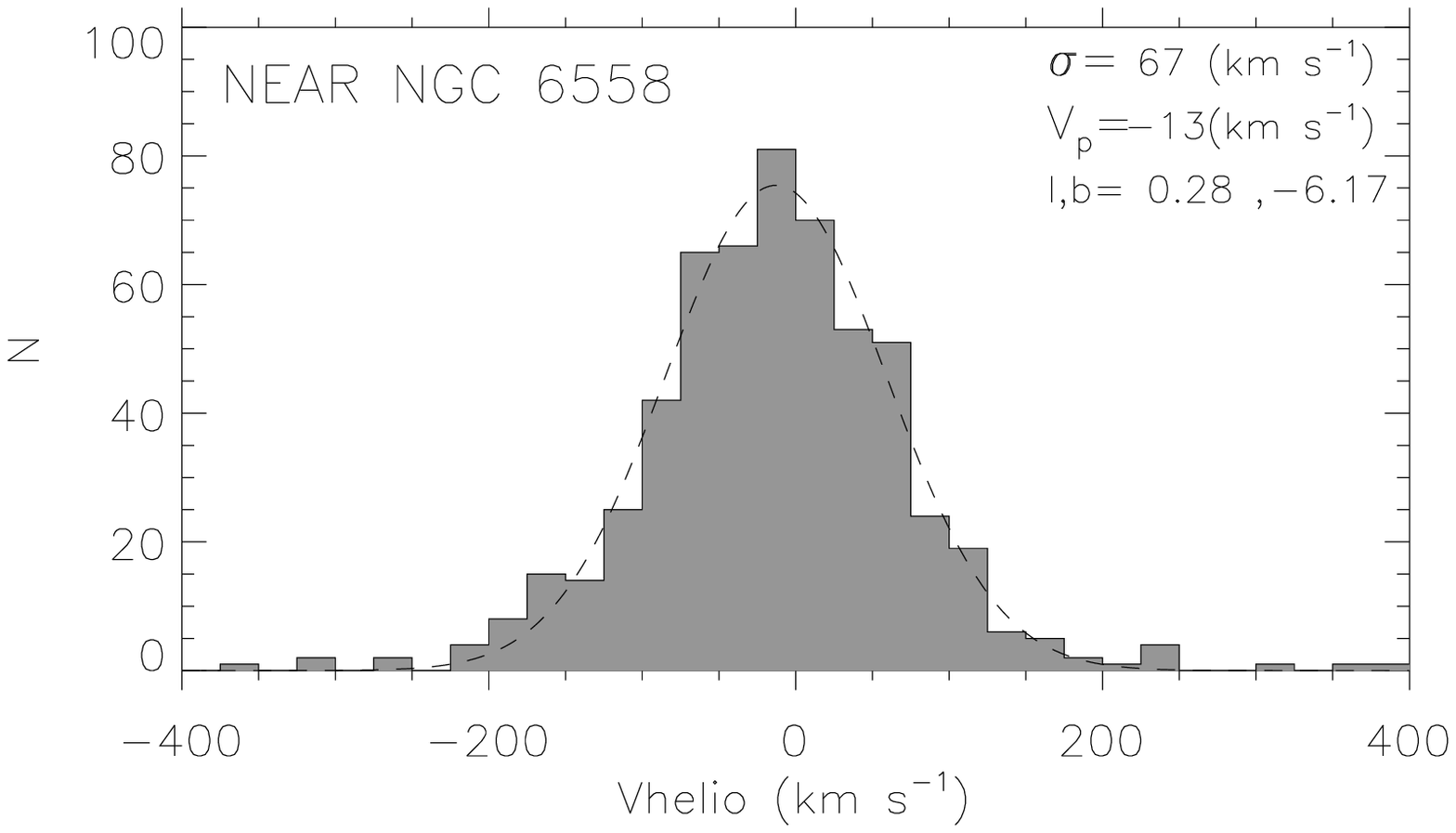}
\includegraphics[width=6.5cm]{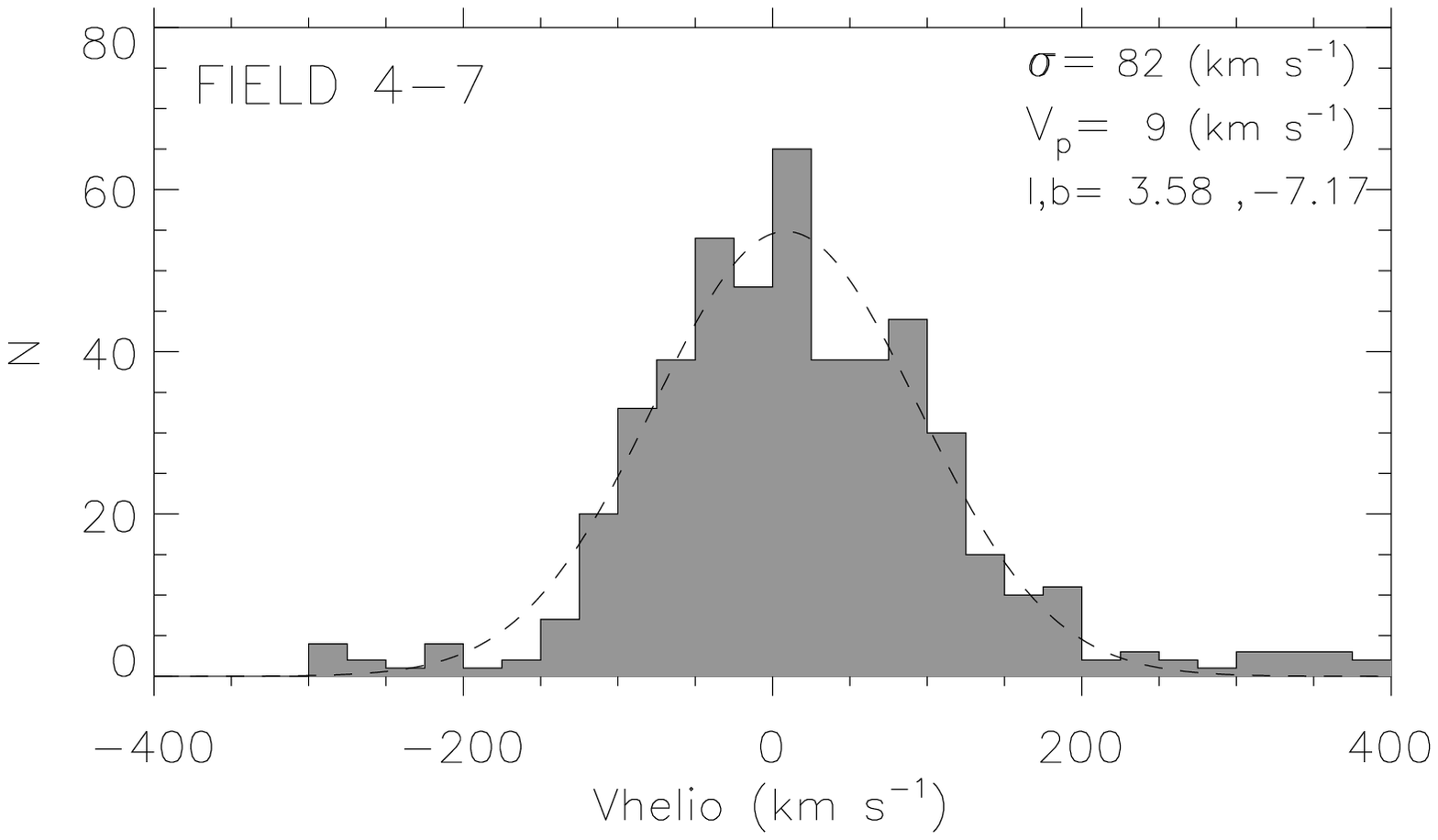}\\
\includegraphics[width=6.5cm]{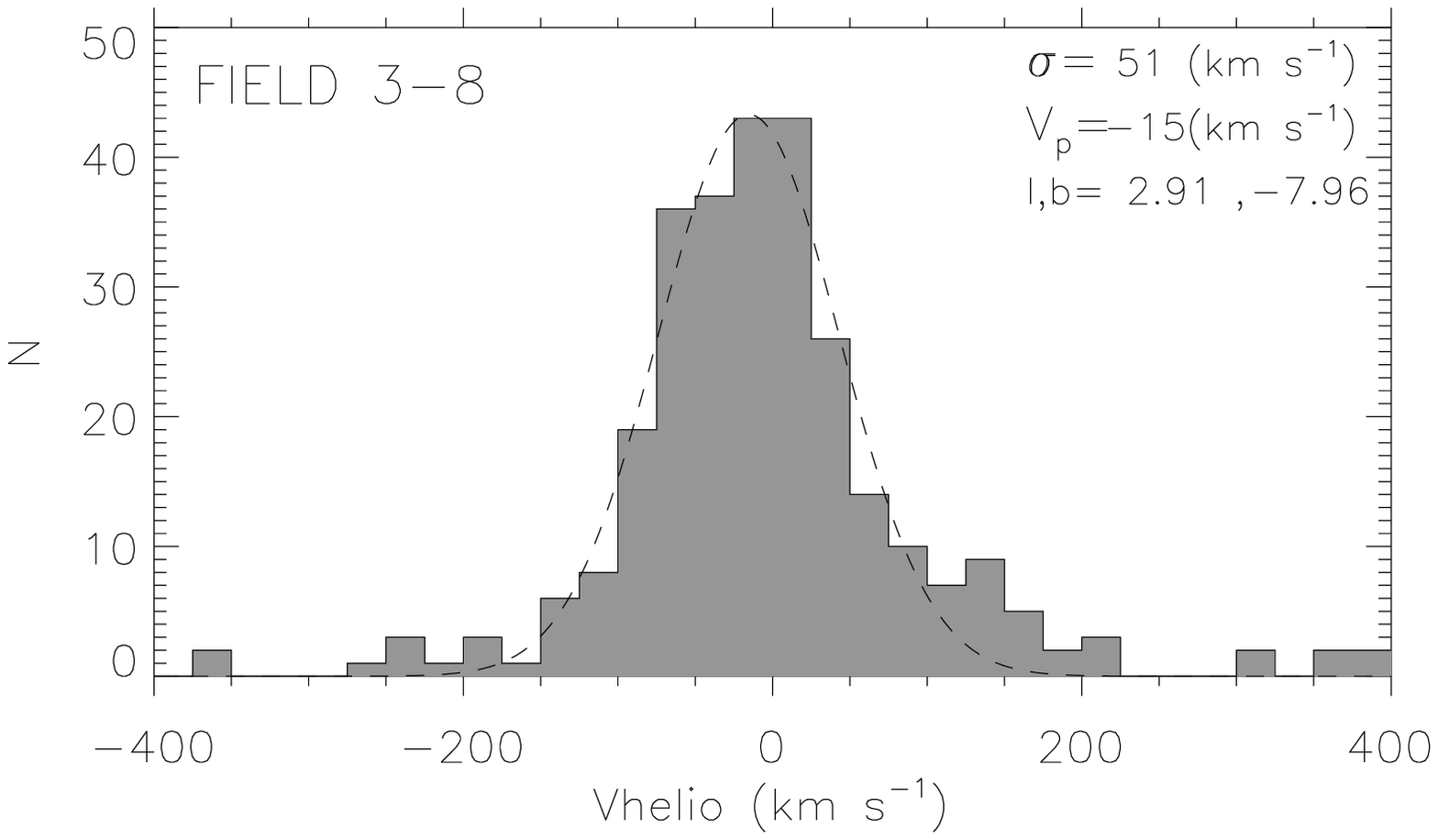}
\includegraphics[width=6.5cm]{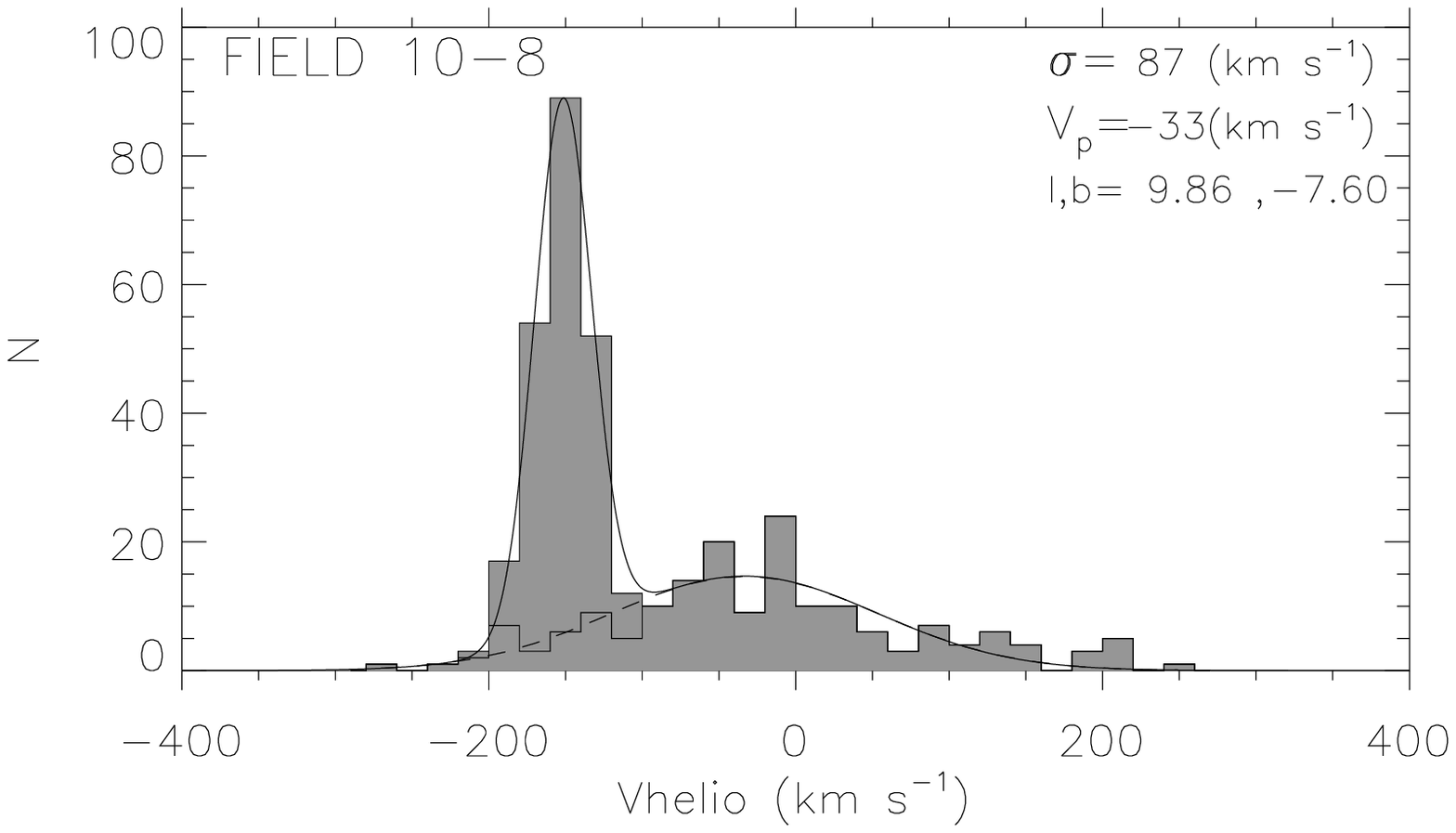}
\caption{Velocity histograms for our six fields. A Gaussian fit has been performed in all
 the cases. Note the contamination by cluster M22 in Field 10-8, which
 appears as the narrow peak at $\sim -150\ km\ s^{-1}$.\label{fig:histovel}} 
\end{figure*}

 Table~\ref{tab:velfit} and Figure~\ref{fig:histovel} show our velocity
 results, where all our velocities are heliocentric and
  have been corrected to  include the zeropoint velocity correction
  and their respective errors. 
 We only selected $\sim 3200$ radial velocities 
 for  these plots, which correspond to stars with velocity errors $\leq 50 \ km\ s^{-1}$.
 Our three minor axis fields Sagittarius-I, Baade's Window and NEAR NGC 6558,
 which target denser parts of the bulge, account for the majority of the results
 ($\sim$2000 radial velocities), while the rest ($\sim$1000 radial velocities)
 are more or less equally distributed between the off-axis fields 
 FIELD 4-7, FIELD 3-8, and FIELD 10-8. 

\begin{table*}
\begin{center}
\caption{Radial Velocity Distribution and Gaussian fit
  parameters for stars with V$_{err} <50$ 
\label{tab:velfit}
}
\begin{minipage}[t]{\textwidth}
\centering
\renewcommand{\footnoterule}{}
\begin{tabular}{ l  c  c  c  c  c }
\hline
\hline
Field       & N$_{Star}$ & $\langle Vr \rangle$ & $Stdev$ &Kurt  & Skew \\ %&  Centre & $\sigma_{G}$\footnote{From histogram Gaussian fit}  
               &      & (km s$^{-1}$)  & (km s$^{-1}$)  &  &  \\ %& (km s$^{-1}$) & (km s$^{-1}$) 
\hline
Sgr-I   &    773  & 5$\pm$4 & 119$\pm$3 & 0.004 & 0.113 \\ %&   4$\pm$4   &  122$\pm$5 
%Sgr-I   (V$_{err} < 30$)   &    655  & 4$\pm$5 & 117$\pm$3 & -0.294 & -0.040 \\ %&   7$\pm$6   &  124$\pm$5
%Sgr-I   &    773  & 5.4$\pm$0.9 & 118.9$\pm$3.0 & 0.004 & 0.113 \\ %&   4$\pm$4   &  122$\pm$5 
%Sgr-I   (V$_{err} < 30$)   &    655  & 4.4$\pm$0.8 & 116.9$\pm$3.2 & -0.294 & -0.040 \\ %&   7$\pm$6   &  124$\pm$5
%\hline
Baade's Window   &    781  &  8$\pm$4 & 119$\pm$3 & 0.057 & 0.258\\ %&   0$\pm$6   & 128$\pm$4
%Baade's Window  (V$_{err} < 30$)    &    621  &  5$\pm$5  &112$\pm$3 & -0.320 & 0.217\\ %&   -3 $\pm$11  & 124$\pm$5
%Baade's Window   &    781  &  8.3 $\pm$ 0.9 & 118.7$\pm$3.0 & 0.057 & 0.258\\ %&   0$\pm$6   & 128$\pm$4
%Baade's Window  (V$_{err} < 30$)    &    621  &  4.5 $\pm$ 0.8  &111.6$\pm$3.2 & -0.320 & 0.217\\ %&   -3 $\pm$11  & 124$\pm$5
%\hline
NEAR NGC 6558   &    563  & -14$\pm$4 &  81$\pm$4  & 2.334 & 0.178\\ %&  -13$\pm$4 & 67$\pm$4 
%NEAR NGC 6558  (V$_{err} < 30$)  &    389  & -17$\pm$4 &  77$\pm$4 & 1.255 & -0.269\\ %&  -14$\pm$4 & 68$\pm$4
%NEAR NGC 6558   &    563  & -13.9$\pm$1.2 &  80.8$\pm$2.4  & 2.334 & 0.178\\ %&  -13$\pm$4 & 67$\pm$4 
%NEAR NGC 6558  (V$_{err} < 30$)  &    389  & -16.5$\pm$1.1 &  76.6$\pm$2.8 & 1.255 & -0.269\\ %&  -14$\pm$4 & 68$\pm$4 
%\hline
FIELD 4-7   &    488  & 19$\pm$5  &  100$\pm$5 & 2.120  & 0.586 \\ %&   9$\pm$5   &  82$\pm$5
%FIELD 4-7  (V$_{err} < 30$)  &    359  & 13$\pm$4  &  82$\pm$5 & 3.430   &  0.767\\ %&   4$\pm$5   &  75$\pm$4
%\hline
%FIELD 4-7   &    488  & 19.2$\pm$1.2  &  100.1$\pm$3.2 & 2.120  & 0.586 \\ %&   9$\pm$5   &  82$\pm$5
%FIELD 4-7  (V$_{err} < 30$)  &    359  & 12.5$\pm$1.1  &  82.4$\pm$3.1 & 3.430   &  0.767\\ %&   4$\pm$5   &  75$\pm$4
%\hline
FIELD 3-8   &    289  & 2$\pm$6    & 106$\pm$8 & 4.290 & 1.135 \\ %&  -15$\pm$4 & 51$\pm$4
%FIELD 3-8  (V$_{err} < 30$)  &    174  & -1$\pm$6   &  83$\pm$9  & 5.368 &  1.680 \\ %&  -19$\pm$6 & 52$\pm$6 
%\hline
%FIELD 3-8   &    289  & 1.6$\pm$1.8    & 105.5$\pm$4.4 & 4.290 & 1.135 \\ %&  -15$\pm$4 & 51$\pm$4
%FIELD 3-8  (V$_{err} < 30$)  &    174  &-1.1$\pm$1.7   &  82.7$\pm$4.5  & 5.368 &  1.680 \\ %&  -19$\pm$6 & 52$\pm$6 
%\hline
FIELD 10-8 \footnote{Distribution of field stars}  &170(365)  & -21$\pm$8  & 98$\pm$5 & -0.190 & 0.311 \\ %&  -33$\pm$8 & 83$\pm$25  
FIELD 10-8 \footnote{Distribution from NGC 6656 stars in FIELD 10-8.} & 228(365)  & -153$\pm$1 & 30$\pm$1 & 1.468  & 0.238 \\ %&  -152$\pm$2  &  24$\pm$2  
%FIELD 10-8\ $^{a}$  (V$_{err} <30$)& 89(187)   & 2$\pm$8  & 70$\pm$6 & 0.384 & 0.561 \\ %&  -18$\pm$21 & 44$\pm$46
%FIELD 10-8\ $^{b}$  (V$_{err} < 30$)& 126(187)  & -148$\pm$2 & 16$\pm$2 & 1.253  & 0.665 \\ %&  -148$\pm$2  &  12$\pm$2 
%
%
%FIELD 10-8 \footnote{Distribution of field stars}  &170(365)  & -21.1$\pm$2.2  & 97.8$\pm$5.3 & -0.190 & 0.311 \\ %&  -33$\pm$8 & 83$\pm$25  
%FIELD 10-8 \footnote{Distribution from NGC 6656 stars in FIELD 10-8.} & 228(365)  & -153.3$\pm$2.2 & 30.0$\pm$2.6 & 1.468  & 0.238 \\ %&  -152$\pm$2  &  24$\pm$2  
%
%FIELD 10-8\ $^{a}$  (V$_{err} <30$)& 89(187)   & 2.1$\pm$2.0  & 70.1$\pm$5.3 & 0.384 & 0.561 \\ %&  -18$\pm$21 & 44$\pm$46
%FIELD 10-8\ $^{b}$  (V$_{err} < 30$)& 126(187)  & -148.4$\pm$2.0 & 15.8$\pm$2.8 & 1.253  & 0.665 \\ %&  -148$\pm$2  &  12$\pm$2 
   
\hline
\end{tabular}
\end{minipage}
\end{center}
\end{table*}

\begin{figure*}
\centering
\includegraphics[width=14cm]{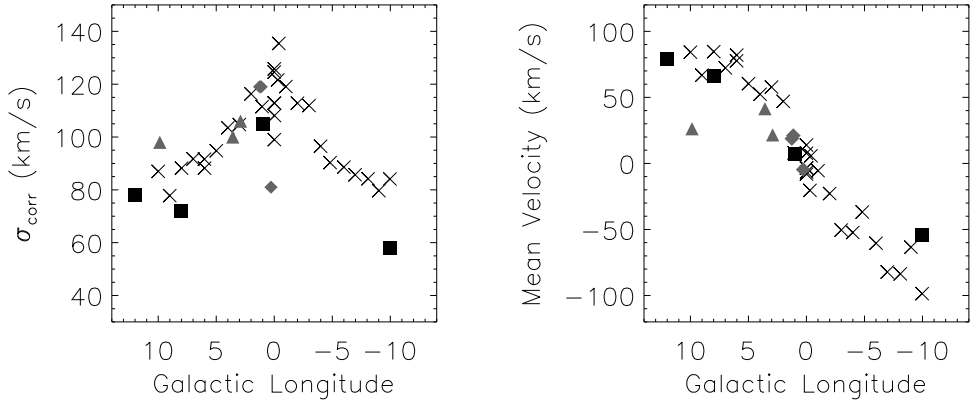} %with velocities from Howard et al. 2008
\caption{ 
 Velocity dispersion profile $\sigma_{corr}$ and rotation curve of
 bulge fields in galactocentric velocity.
The plot includes several samples: K-giants by 
 Minniti 1996 (squares), M-giants by \cite{howard2008} (crosses), and our fields
 (grey diamonds for minor axis fields and grey triangles for off-axis fields). In some fields
 we were unable to separate clearly the bulge and disk
 foreground members (see KR02). This contamination problem was more 
 significant in the fields at positive longitudes and field NEAR NGC 6558.
\label{fig:rotgal}} 
\end{figure*}

\begin{figure*}
\begin{center}
\includegraphics[width=6.5cm]{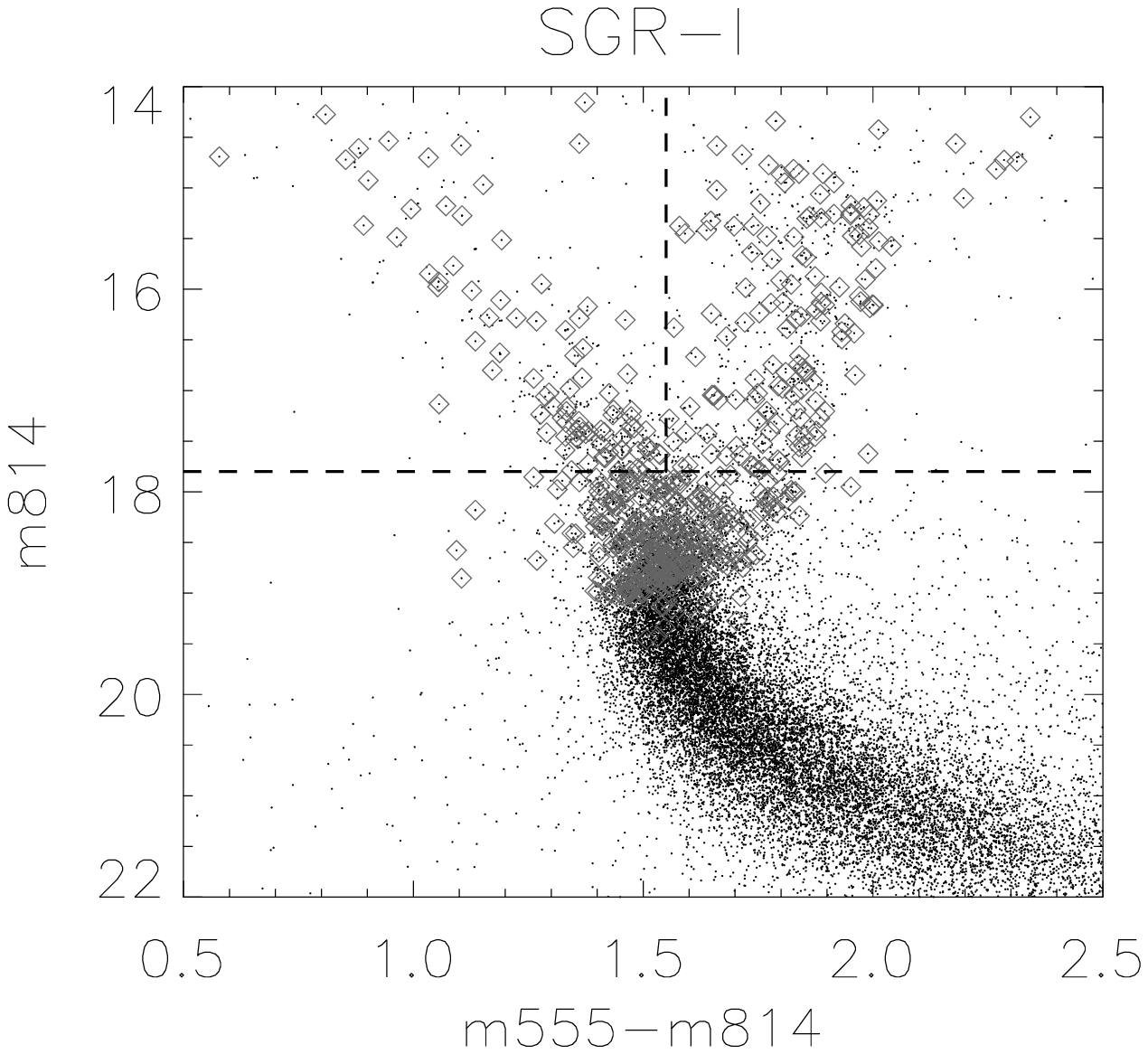}
\includegraphics[width=6.5cm]{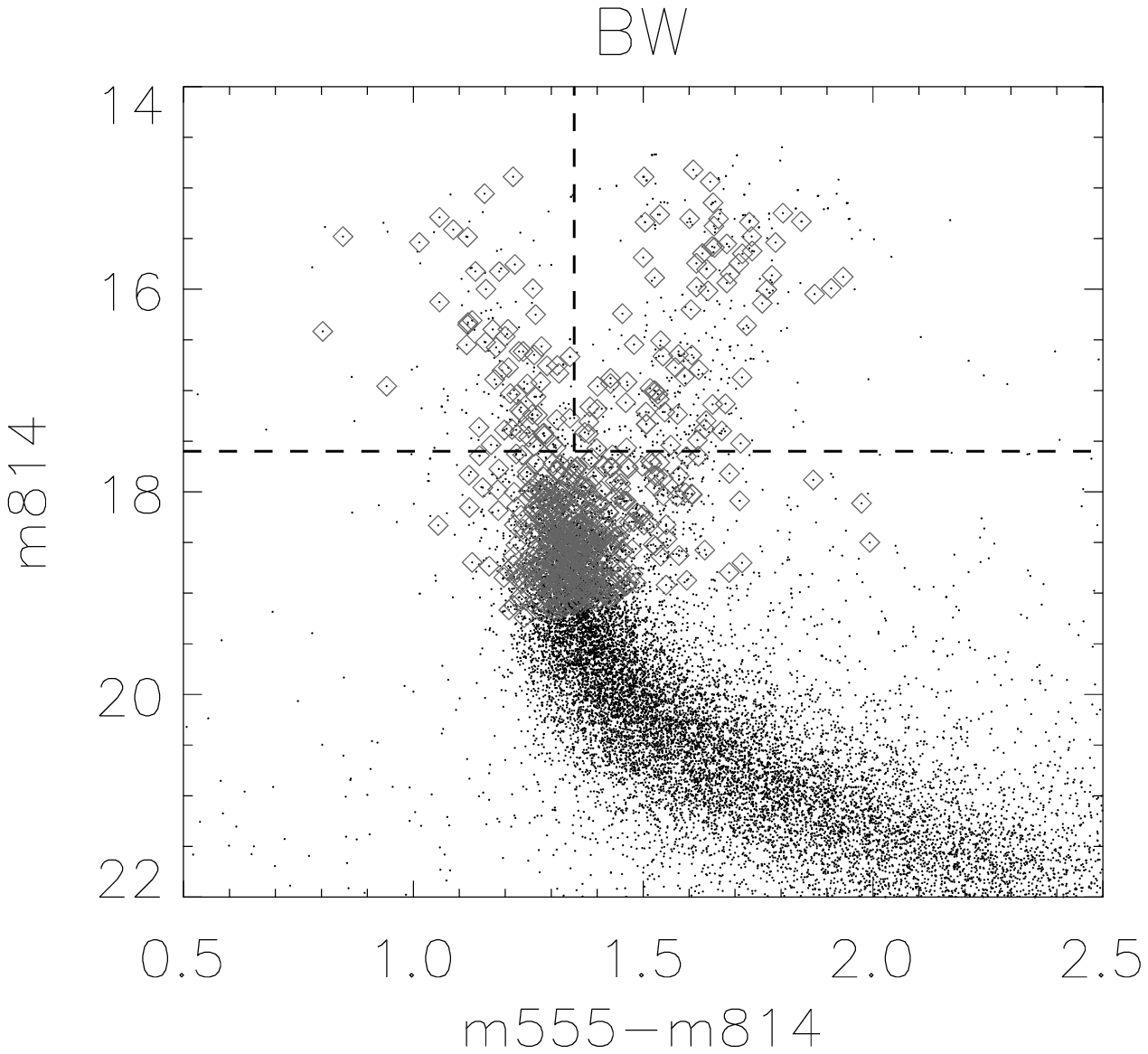}  
\includegraphics[width=6.5cm]{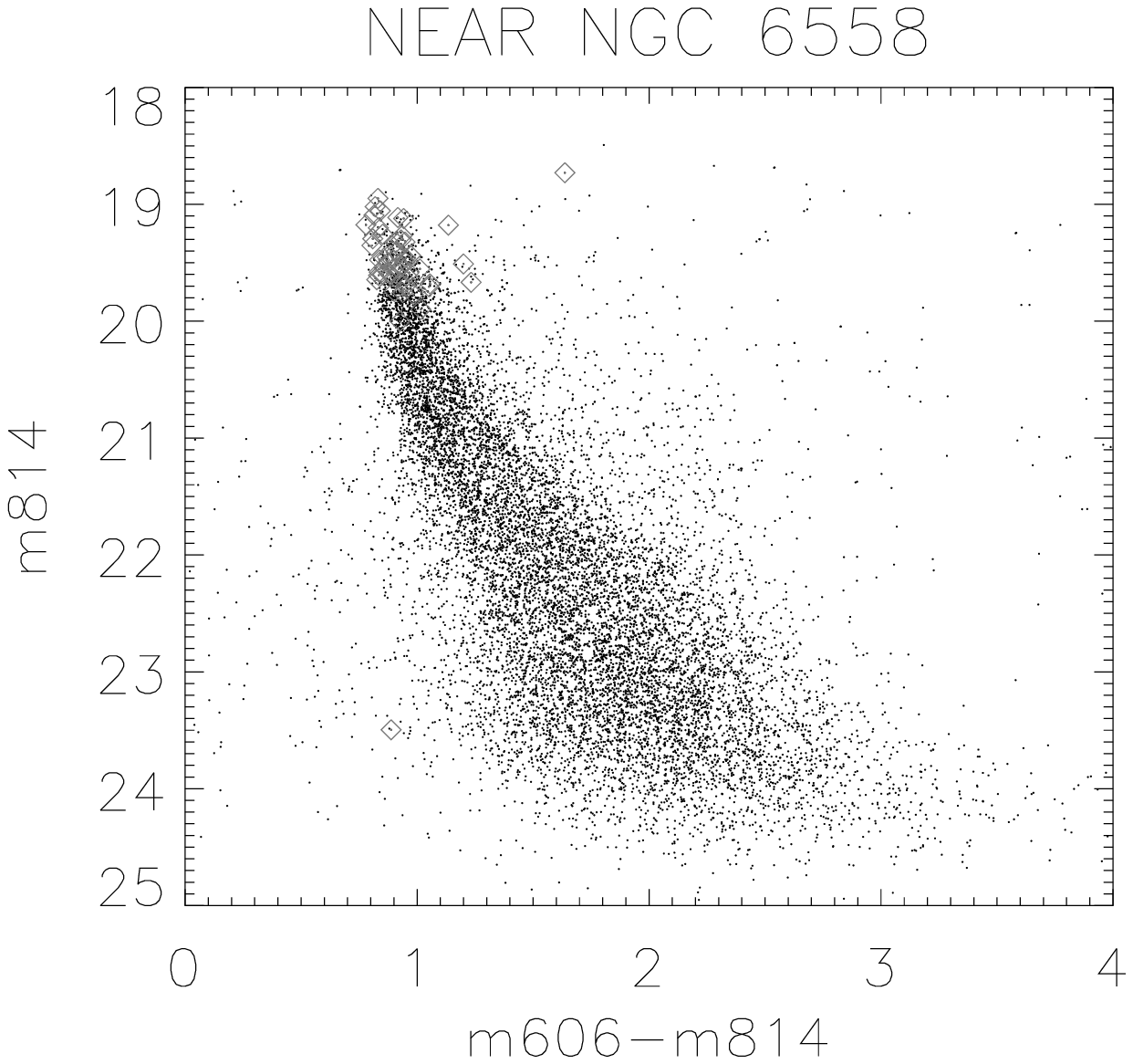}
\caption{ CMD  for the three minor-axis fields. IFU detections (open squares) have 
 been highlighted over 
  HST detections in each field. Dashed lines in   
 Sagittarius-I and Baade's Window fields represent the limits of the
 different regions in which we have divided each CMD, blue-end of the
 main sequence (blue-MS; top left), Reg Giant Branch (RGB; top right) and
 turn-off (bottom). Consequently, we have   
 excluded the blue-end of the main sequence from the 
 bulge analysis due to its proper motions; these are consistent with a
 population dominated by foreground stars rotating in front of the bulge. 
\label{fig:CMD}}
\end{center}
\end{figure*}
  
 The presence of the cluster  NGC 6656 (also known as M22) 
 on Field 10-8 has been used to check the accuracy of our radial velocity 
 calibrations. Mean radial velocity and dispersion of M22 are well known  
 and have been checked against our results; \cite{peterson94} (1994) reported 
 a mean velocity of $-148.8\ \pm$ 0.8 $km\ s^{-1}$ and an intrinsic velocity 
 dispersion of 6.6 $\pm$ 0.8 $km\ s^{-1}$. The former value is in clear agreement 
 with our reported value of 
 -153$\pm$1 $km\ s^{-1}$. %  for the center of the Gaussian fit. 
  Furthermore, this agreement improves if we limit
 our sample only to stars with velocity errors below 30 $km\ s^{-1}$
 (126 stars), with a 
 mean velocity of -148$\pm$2. %  for the Gaussian center of the fit.   

 Similarly, Baade's Window velocities are well documented, 
 and also can be used to asses the accuracy of our results. Hence, we found that 
Baade's Window and  
Sagittarius-I have high velocity dispersions in our results, 
 as Table~\ref{tab:velfit} shows. 
 If we constrain once again the sample only to stars with velocity errors below 
 30 $km\ s^{-1}$ (621 stars), the dispersion for Baade's Window drops to 
 112$\pm$3 $km\ s^{-1}$ with a mean velocity of 5$\pm$5 $km\ s^{-1}$. 
 This measurement shows an excellent agreement
 with other studies:  
 Babusiaux et al. (2010): $\sigma=111\pm 4\ km\ s^{-1}$, $\langle Vr \rangle= 9\pm6\ km\ s^{-1}$,  
 Rangwala et al. (2009):  $\sigma=112\pm 3\ km\ s^{-1}$, $\langle Vr \rangle= 9\pm6\ km\ s^{-1}$,
 Rich et al. (2007): $\sigma=110\pm9\ km\ s^{-1}$, $\langle Vr \rangle= -1\pm13\ km\ s^{-1}$, 
 Terndrup et al. (1995): $\sigma=110\pm10\ km\ s^{-1}$, $\langle Vr \rangle= -8\pm6\ km\ s^{-1}$, 
 Sharples et al. (1990): $\sigma=113_{-5}^{+6}\ km\ s^{-1}$, $\langle Vr \rangle= 4\pm8\ km\ s^{-1}$.     
 
 Kurtosis in the two minor-axis fields, Baade's Window and Sagittarius-I, are consistent with 
 normal distributions. Conversely, field NEAR NGC 6558 shows a pointy distribution which suggests
 disk contamination; the same is repeated for Field 4-7 and Field 3-8.  
 Galactic rotation and velocity dispersion curves from previous studies (Figure~\ref{fig:rotgal}) and based on K-giants and M-giants 
 at several longitudes (Minniti 1996, and Rich et al. 2007, respectively) 
 also show consistency with most of our fields.
 On the contrary, we found inconsistency in the velocity dispersion and Galactic rotation curve for Field 
 10-8 and NEAR NGC 6558 respectively.
 NEAR NGC 6558 shows a small dispersion compared with the other two minor-axis fields, 
 Sagittarius-I and Baade's Window, 
 and it is under the curve of velocity dispersion of the galaxy. Nevertheless,
 NEAR NGC 6558 agrees with the bulge rotation curve.   
 In our off-axis fields we found that Field 4-7 and Field 3-8 seem to show 
 a reasonable agreement with the bulge rotation curve.  
  FIELD 10-8 on the other hand, does not show a good agreement. 
 This apparent disagreement in the latter field and NEAR NGC 6558 might be caused by 
 strong contamination by non-bulge stars (mainly cluster stars from M22 and NGC 6558
 respectively), or by poor statistics. 
 Other authors have previously explored contamination rates in Field 10-8.
 (Minniti et al. 1996)

 A key aim of our work is to derive space motions for a large sample 
 of bulge stars, combining radial radial velocities and proper motions.
 As shown in KR02, main-sequence stars 
 brighter than the turn-off 
 show a proper motion drift consistent with a foreground disk population rotating in front
 of the bulge, whereas red giants show kinematics representative of the bulge 
 population as a whole. We attempt to remove foreground stars via cuts in the CMD
 as it is shown in Figure~\ref{fig:CMD}.
\begin{table*}
\begin{center}
\begin{minipage}{\textwidth}
\renewcommand{\footnoterule}{}
\caption{Velocity ellipsoid parameters for Galactic minor-axis fields
  Baade's Window and Sagittarius-I. The first
  three rows correspond to the combination of turnoff and RGB
  populations which should be dominated by bulge populations (Kuijken
  \& Rich 2002) and according to the cuts in Figure \ref{fig:CMD}.
   Last 6 rows are the parameters for the velocity
  ellipsoids in each field divided by populations, turnoff, RGB and
  blue MS.
}
\label{table:velellip}      % is used to refer this table in the text 
%\centering 
\begin{center}
\begin{tabular}{ l  c  c  c  c  c  c  c  c } 
\hline
\hline
Field            & $\sigma_r$   &   $\sigma_l$  & $\sigma_b$ &
r$_S$\footnote{Spearman's correlation coefficient between $\mu_l$ and $Vr$} & Prob(r$_S$) \footnote{Prob(r$_S$) corresponds to the significance of the correlation r$_S$} & $l_v$\footnote{Vertex angle} & N\footnote{Total number of stars selected} & N$_{rej}$\footnote{Number of rejected stars during clipping procedure}  \\
             & km s$^{-1}$ &  km s$^{-1}$ & km s$^{-1}$ &  &   & ($^{\circ}$) &  &  \\   
\hline
%
%Sgr-I        & 121.4$\pm$3.0 & 117.8$\pm$3.3 &  102.9$\pm$2.9  & -0.181 & 2e-5 & -39$\pm$7 & 560 & 9 \\        
%BW           & 116.8$\pm$3.0 & 116.9$\pm$3.3 & 103.7$\pm$4.9 & -0.285 & 5e-13 & -43$\pm$4 & 616 & 10 \\ 
%NEAR NGC 6558     & 67.9$\pm$3.5 & 93.4$\pm$4.1  & 90.2$\pm$4.9  & -0.108 & 0.095  & -17$\pm$7  & 242  & 12 \\
% & & & & & & & & \\
%Sgr-I (turn-off) & 125.5$\pm$3.7 & 118.2$\pm$3.9 &  104.3$\pm$3.7  & -0.170 & 7e-4 & -34$\pm$6 & 397 & 8 \\
%Sgr-I (RGB)      & 110.7$\pm$6.0 & 114.8$\pm$5.7 &  99.0$\pm$5.3  & -0.200 & 0.010 & -42$\pm$9 & 163 & 1 \\
%Sgr-I (blue-MS)  & 88.5$\pm$8.9 & 114.1$\pm$8.9 &  99.2$\pm$8.3  & -0.030 & 0.782 & -14$\pm$11 & 86 & 3 \\
% & & & & & & & & \\
%BW(turn-off) & 119.5$\pm$3.0 & 117.5$\pm$3.1 & 104.4$\pm$5.8  & -0.292 & 1e-11 & -41$\pm$6 & 512 & 8 \\
%BW(RGB)      & 108.2$\pm$6.9 & 114.3$\pm$7.7  & 100.9$\pm$6.2  & -0.241 & 0.014 &  -40$\pm$13 & 104 & 1 \\
%BW(blue-MS) & 94.1$\pm$8.1 & 131.9$\pm$9.8 & 90.9$\pm$9.0 & -0.149 & 0.211 & -21$\pm$7 & 72 & 1 \\
%
Sgr-I        & 121$\pm$3 & 118$\pm$3 &  103$\pm$3  & -0.181 & 2e-5 & -39$\pm$7 & 560 & 9 \\        
BW           & 117$\pm$3 & 117$\pm$3 & 104$\pm$5 & -0.285 & 5e-13 & -43$\pm$4 & 616 & 10 \\ 
NEAR NGC 6558     & 68$\pm$4 & 93$\pm$4  & 90$\pm$5  & -0.108 & 0.095  & -17$\pm$7  & 242  & 12 \\
 & & & & & & & & \\
Sgr-I (turn-off) & 125$\pm$4 & 118$\pm$4 &  104$\pm$4  & -0.170 & 7e-4 & -34$\pm$6 & 397 & 8 \\
Sgr-I (RGB)      & 111$\pm$6 & 115$\pm$6 &  99$\pm$5  & -0.200 & 0.010 & -42$\pm$9 & 163 & 1 \\
Sgr-I (blue-MS)  & 88$\pm$9 & 114$\pm$9 &  99$\pm$8  & -0.030 & 0.782 & -14$\pm$11 & 86 & 3 \\
 & & & & & & & & \\
BW(turn-off) & 119$\pm$3 & 117$\pm$3 & 104$\pm$6  & -0.292 & 1e-11 & -41$\pm$6 & 512 & 8 \\
BW(RGB)        & 108$\pm$7 & 114$\pm$8 & 101$\pm$6  & -0.241 & 0.014 &  -40$\pm$13 & 104 & 1 \\
BW(blue-MS) & 94$\pm$8   & 132$\pm$10 & 91$\pm$9  & -0.149 & 0.211 & -21$\pm$7 & 72 & 1 \\
\hline
\end{tabular}
\end{center}
\end{minipage}
\end{center}
\end{table*}
 Unfortunately, no such cuts could be applied to
 the NEAR NGC 6558 field, for which no suitable archival first epoch images 
 are available, giant and turn-off stars appear saturated on the respective CMD. 
 Hence, proper motions 
 are only available for fainter stars in this field, where we expect disk contamination 
 to be less pronounced at the latitude of this field when
 compared with the fields SGR-I and Baade's Window.
 Our three minor-axis fields, whose results we analyze here, 
 SGR-I, Baade's window and NEAR NGC 6558 lie close to
 the Galactic minor axis. An axisymmetric bulge should produce at those
 longitudes 
 velocity ellipsoids aligned with the line of sight. Therefore,
 any deviation of that alignment in the velocity ellipsoid or 
 vertex deviation
 is a clear signature of a non-axisymmetric bulge. 
\begin{figure*}
\centering
\includegraphics[width=5.5cm]{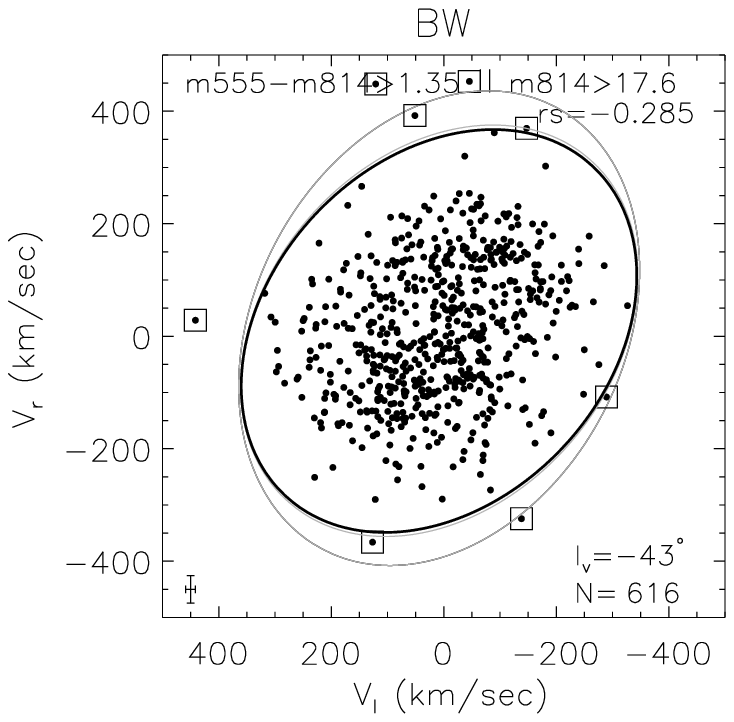}
\includegraphics[width=5.5cm]{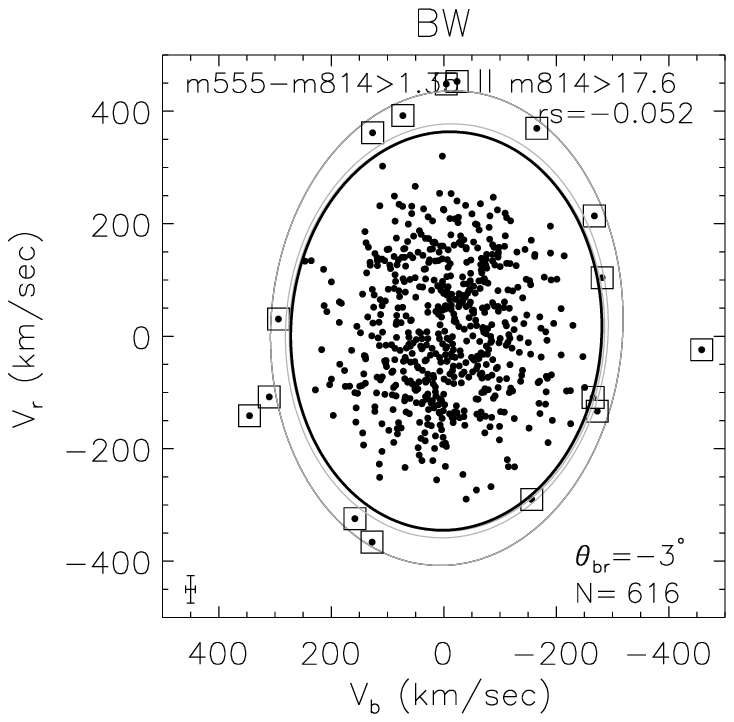}
\includegraphics[width=5.5cm]{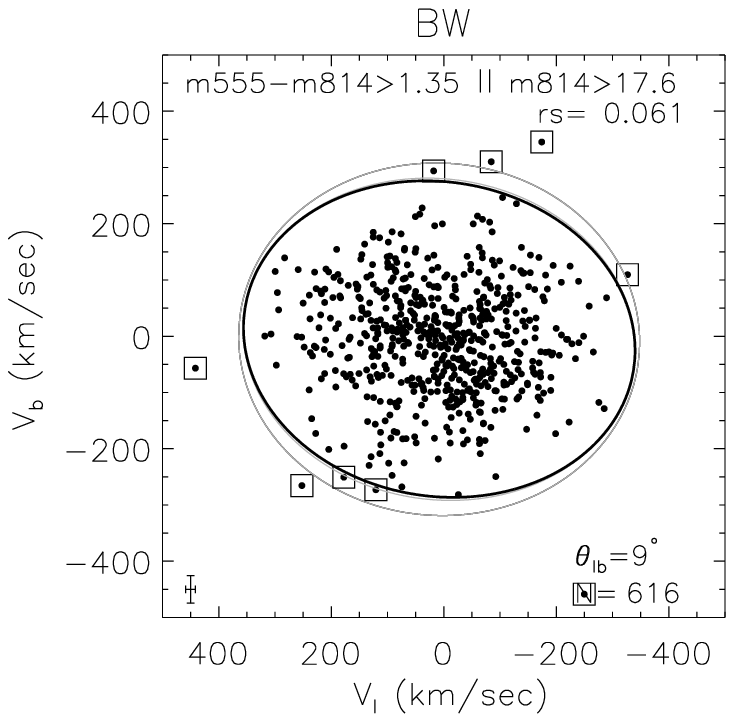}
\includegraphics[width=5.5cm]{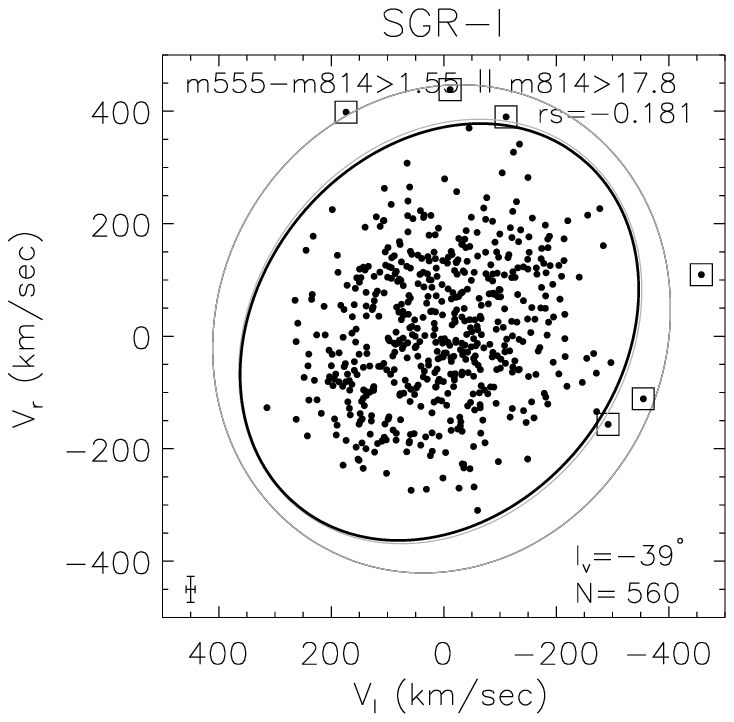}
\includegraphics[width=5.5cm]{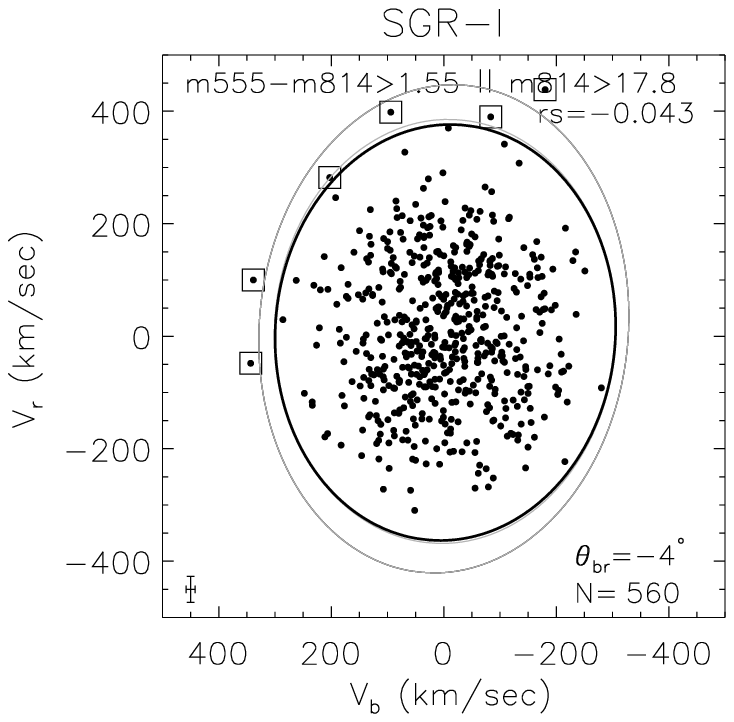}
\includegraphics[width=5.5cm]{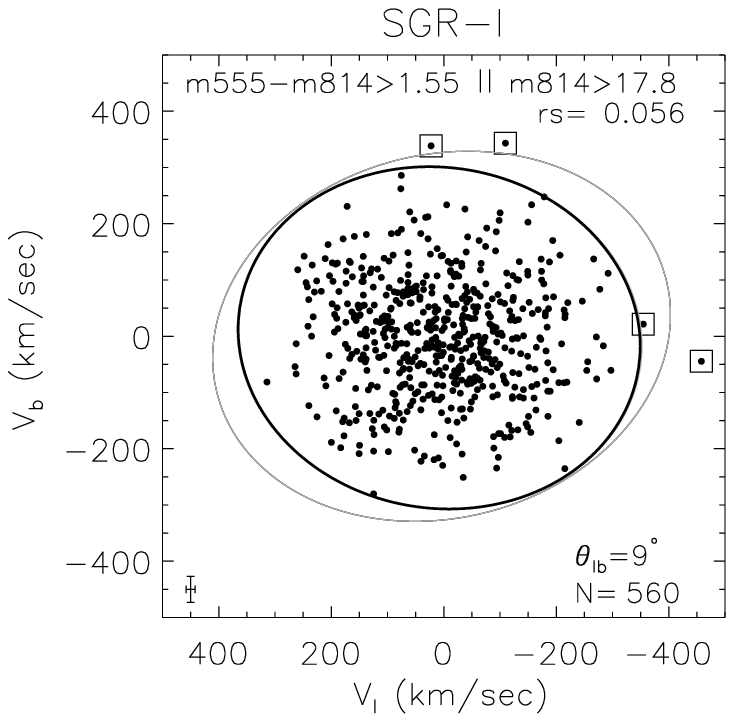}
\includegraphics[width=5.5cm]{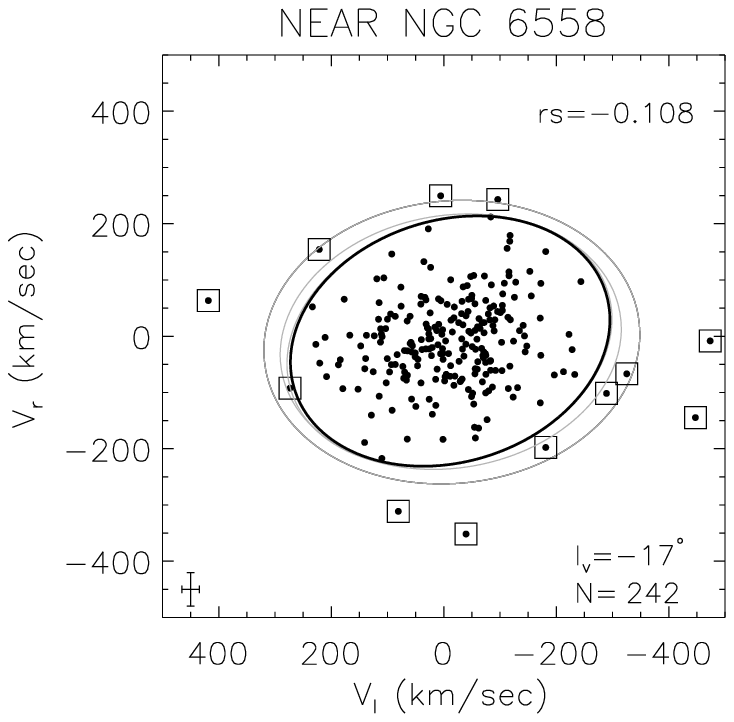}
\includegraphics[width=5.5cm]{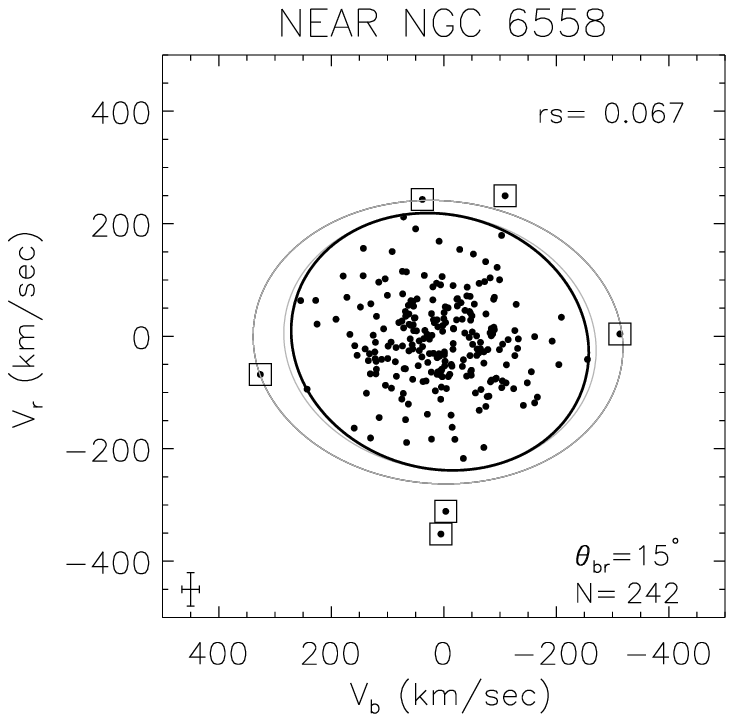}
\includegraphics[width=5.5cm]{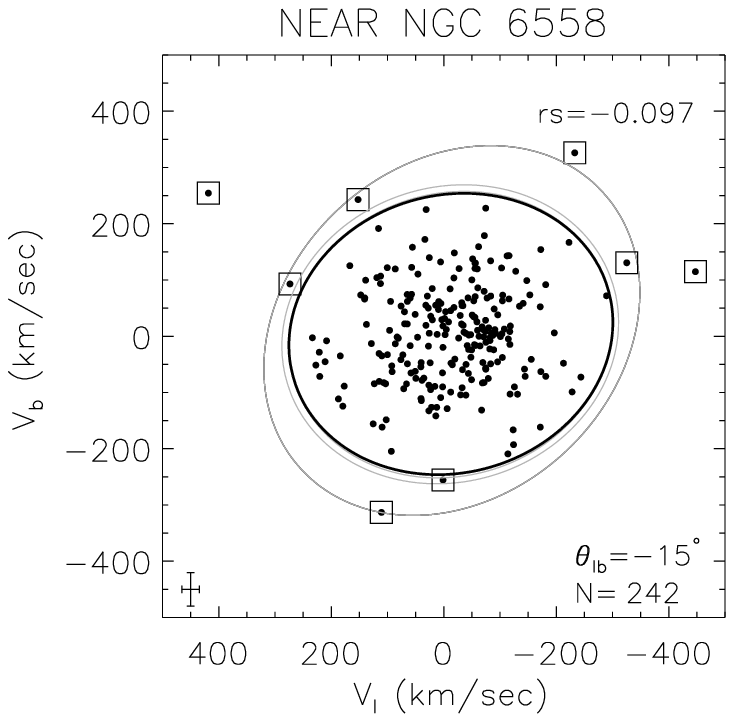}
\caption{Velocity ellipsoids for our minor-axis fields. 
 Correlation between $V_l$ and $V_r$
 can be related with bulge triaxility. The color-magnitude selection has been the 
 same shown in Figure~\ref{fig:CMD} and numbers of the cuts are 
 shown in every plot when present (e.g. for Sagittarius-I the selection is 
 [m555-m814$>$1.55 or m814$>$17.8], which includes turnoff and RGB stars). 
 An iterative clipping 
 algorithm to reject stars outside 3.0 $\sigma$ of the velocity ellipsoid 
 was used to rid the sample of stars with a kinematic behavior different
 than the majority. Each figure includes all the velocity ellipsoids to
 illustrate the convergence of the method. Enclosed points are those 
 rejected for the final velocity ellipsoid for which the parameters found 
 have been included in Table~\ref{table:velellip}.
 In addition, for the final selection has been calculated in each case  the
   Spearman's correlation coefficient $r_S$ and the vertex angle $l_v$
   when possible.
 \label{fig:velellip} }   
  
\end{figure*}
Specifically, a vertex
 deviation should appear in the correlation of transverse proper motion 
 $\mu_l$ and the radial velocity Vr. Although this correlation will be
 affected by line of sight projection and bulge shape, a significant correlation 
 should be a robust bar indicator. We therefore computed 
 the dispersion tensor $\sigma_{ij}^2$. Eigenvalues and eigenvectors 
 of the dispersion 
 tensor correspond to the axis ratio and direction of the axis of the 
 ellipsoid, while the center is given by the velocity first moments.
  These calculations have been summarized in Table~\ref{table:velellip} 
 and plotted in Figure~\ref{fig:velellip}. 
 Values $\sigma_r$, $\sigma_l$ and $\sigma_b$ 
 are the eigenvalues of the velocity ellipsoid and r$_S$ is the Spearman's 
 correlation coefficient.

\begin{figure*}
\centering
\includegraphics[width=6.0cm]{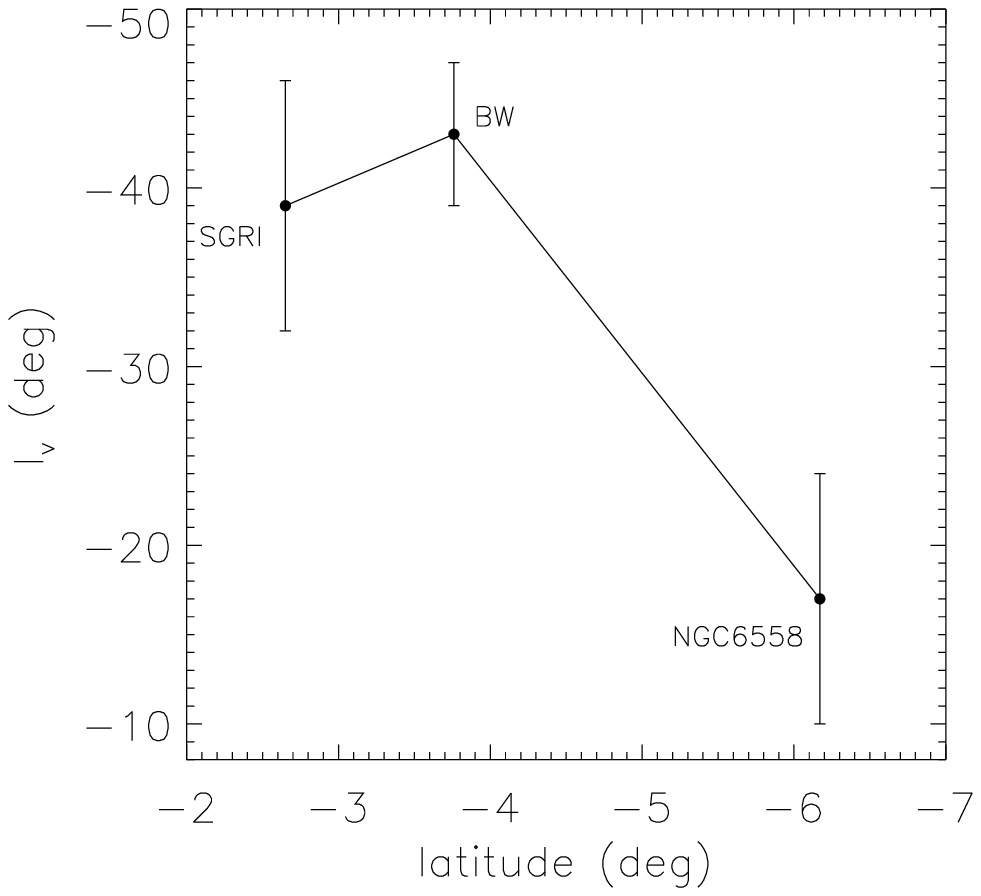}
\caption{ Vertex deviation $l_v$ as a function of the Galactic
  latitude $b$ for the three minor axis fields on this project
  Sagittarius-I, Baade´s Window and NEAR NGC 6558. 
 \label{fig:vertexvslat}}   
  
\end{figure*}

 The stars shown in Figure~\ref{fig:velellip} were selected in the CMD with
 the same cuts as before. An iterative clipping algorithm was used to derive 
 robust dispersions.
 Figure~\ref{fig:velellip} includes all the iterative 
 velocity ellipsoids, and
 points within squares are the stars rejected during the process. 
 The values in Table 
 ~\ref{table:velellip} are those found in the last velocity ellipsoid.    
  We find significant correlations in
 our minor-axis fields; the vertex deviations are virtually 
 the same for the Baade's Window and Sagittarius-I, -43$^{\circ} \pm$4$^{\circ}$ 
 and -39$^{\circ}\pm$7$^{\circ}$ respectively, and -17$^{\circ}\pm$7$^{\circ}$ in NEAR NGC 6558. 
 Similarly the Spearman correlation coefficient r$_S$ shows significant correlations
 for Baade's Window and Sagittarius-I, the significance (Prob(r$_S$)) shows a probability over 99.9\% 
 that correlations are real in Sagittarius-I and Baade's Window. The latter probability decreases to
 90\% in the case of NEAR NGC 6558 in agreement with the vertex angle results.
 Thus, we have a clear trend of a decreasing vertex deviation towards the 
 Galactic plane in the minor-axis which is shown in Figure
 \ref{fig:vertexvslat}. These bar signatures even
 at latitudes of $b \sim -6^{\circ}$ agree with Shen et
 al. (2010) N-body simulations of the Galactic bar, where they find 
 bar signatures at $b \sim -8^{\circ}$.
 Correlations in the other components of the velocity ellipsoid were also explored. 
 In our sample, the equivalent of the vertex deviation ($l_v=\theta_{lr}$) for 
 the $br$ and $lb$ components ($\theta_{br}$ and $\theta_{lb}$ respectively) showed no significant 
 correlations for the three minor axis fields. 

 Even though our selection lacks metallicities, which have
 been mentioned as a relevant parameter to separate between disk and 
 bulge populations (Minniti 1993; Zhao et al. 1994), the agreement with
 such samples with well established metallicities would support our selection criterion.
 Zhao et al. (1994) analyzed a  subsample
 of 62 stars with metallicities, radial velocities 
 (Rich 1988, 1990) and proper motions 
 from the original sample of 400 K and M giants by Spaenhauer et al. (1992).
 In the case of Zhao et al (1994), only a vertex deviation was
 found in the metal rich population of his small sample (39 stars). 
  More recently, the original proper motions by Spaenhauer et al. (1992) have been
 complemented with radial velocity and metallicity measurements 
 (Terndrup et al. 1995, Sadler et al. 1996) 
 allowing us to have $\sim$300 stars with well defined 3-D kinematics 
 and abundances.
 The result of this 
 increased sample has been consistent with the 
 preliminary Zhao results (\cite{soto}) showing a 
 significant vertex deviation 
 only for the metal rich stars. Moreover, the sample of K giants shows a 
 remarkable agreement in $l_v$ with our own sample of turn-off and main sequence stars,
 where the angles found are -34$^{\circ} \pm$7$^{\circ}$ and -43$^{\circ} \pm$4$^{\circ}$
 respectively. A similar agreement is found for the Babusiaux et al. (2010) sample, 
 with a vertex deviation of -36$^{\circ} \pm$5$^{\circ}$ in the metal rich population.
 Furthermore, our reported anisotropies in Baade's Window and
  Sagittarius-I ($\sigma_b/\sigma_r=$0.89$\pm$0.05 and
  $\sigma_b/\sigma_r=$0.85$\pm$0.03 respectively) 
 also suggest the flattening of the bar population which is consistent
 with Babusiaux et al. (2010) metal rich population ($\sigma_b/\sigma_r=$0.78$\pm$0.08).   
 In the same way, we found $\sigma_r \geq \sigma_l>\sigma_b$ for our two fields
 Baade's Window and Sagittarius-I.

\begin{figure*}
\centering

\includegraphics[width=5.5cm]{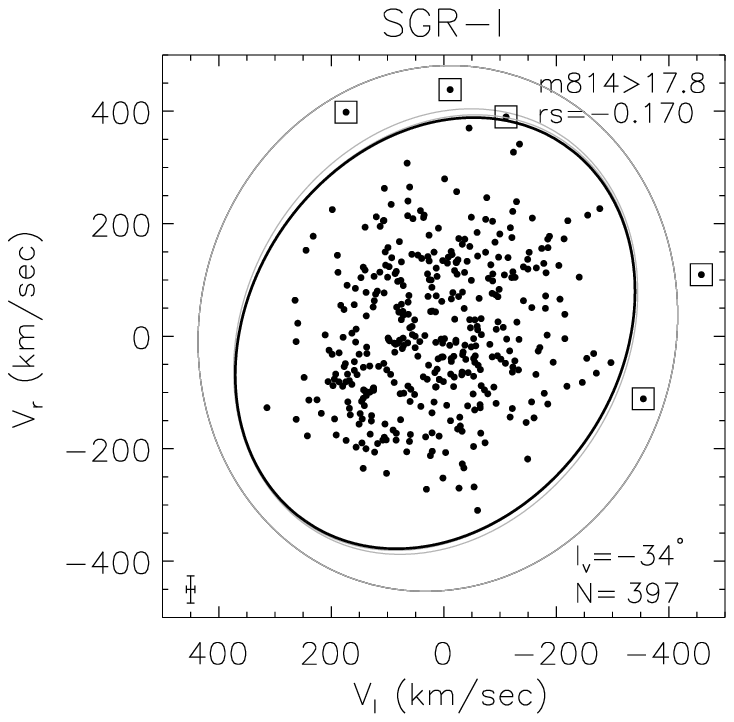}
\includegraphics[width=5.5cm]{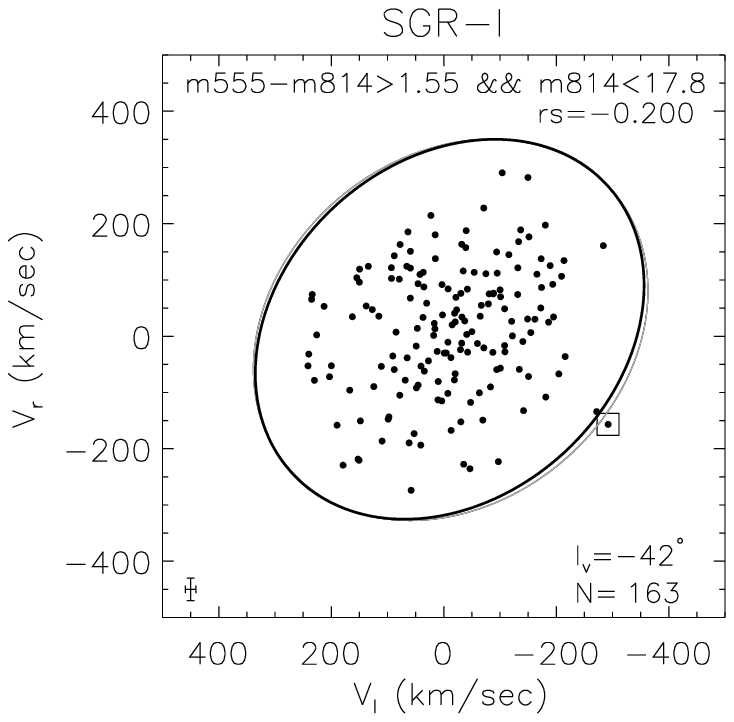}
\includegraphics[width=5.5cm]{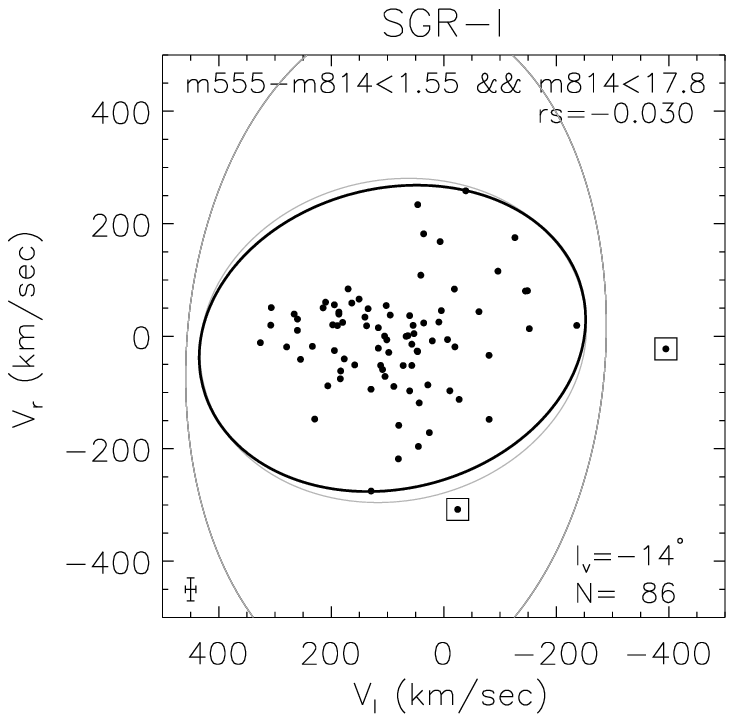}\\
\includegraphics[width=5.5cm]{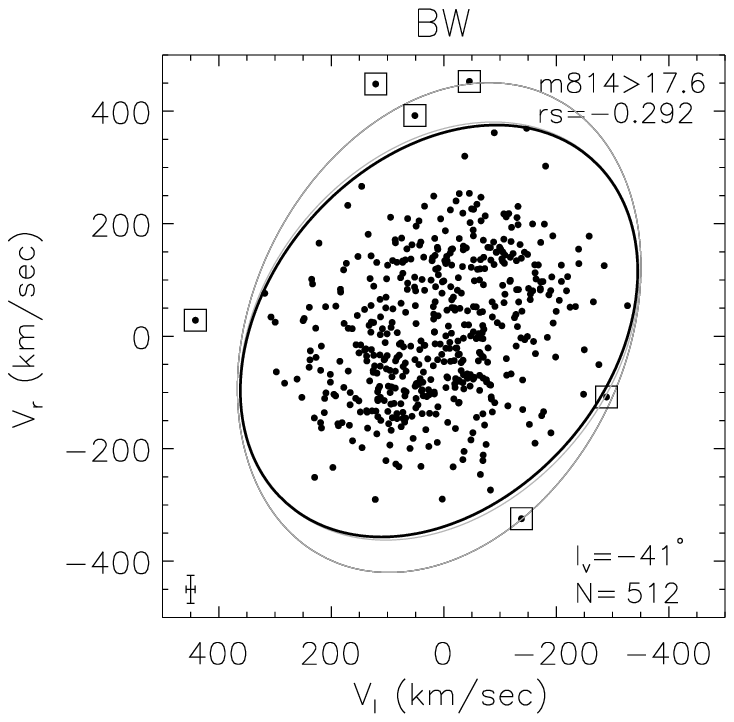}
\includegraphics[width=5.5cm]{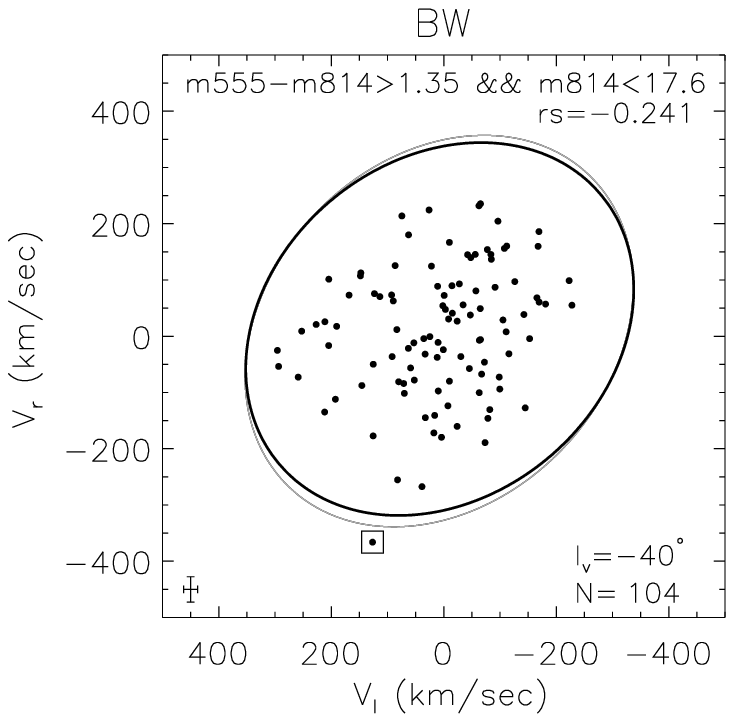}
\includegraphics[width=5.5cm]{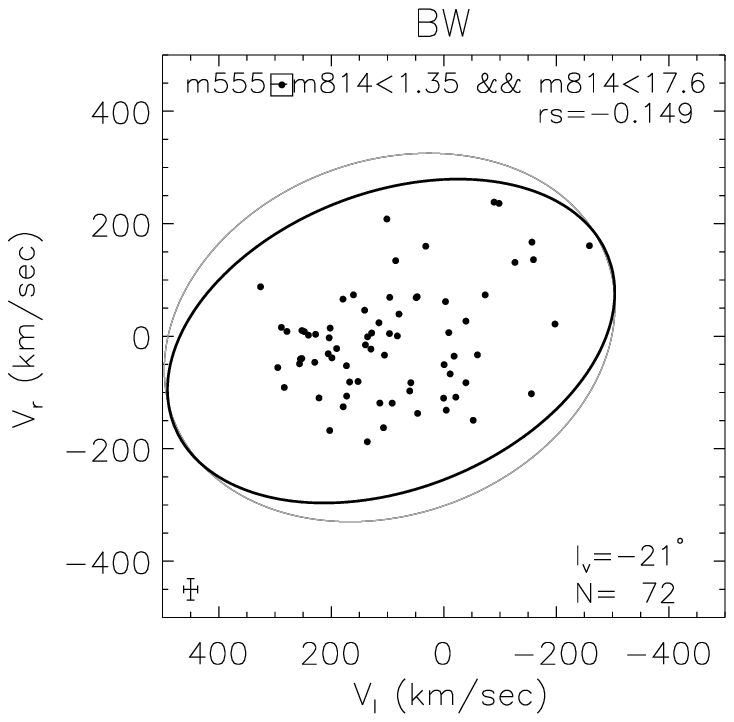}
\caption{Baade's Window and Sagittarius-I velocity ellipsoids for radial
 velocities ($V_r$) and  transverse proper motion ($V_l$), divided by
  population. The region selected is indicated in each plot, and
  follows the same limits shown in figure \ref{fig:CMD}, which correspond
 (from left to right) to turn-off, RGB and blue-end of the main sequence.
\label{fig:velellip2} }   
  
\end{figure*}

 As an additional exercise, we have explored the velocity ellipsoids for 
 Sagittarius-I and BW, dividing the sample in turnoff, blue-end of the 
 main sequence (blue-MS) and red giant branch
 (RGB). In order to separate into populations,
 we have used the same limits in color and magnitudes as before to exclude 
 disk stars. Table~\ref{table:velellip} and 
 Figure~\ref{fig:velellip2} show these ellipsoids. Turnoff selection from 
 the color-magnitude cuts 
 concentrates most of the stars in both samples, repeating the distribution observed
 in Figure ~\ref{fig:velellip}. 
 Similarly, RGB velocity ellipsoids follow the same trend
 as expected by KR02 binned CMD. At the same time, bright blue main sequence stars show no 
 significant correlation in both fields, where the velocity ellipsoids have converged to 
 vertex angles of $\sim$10$^{\circ}$ as would be expected by a
 population dominated by foreground stars rotating in front 
 of the bulge. Moreover, the blue-MS velocity ellipsoids show
 agreement with Babusiaux et al. (2010) metal poor population, $\sigma_l$ is higher
 than the other components ( $\sigma_r \simeq \sigma_b$ in Baade's
 Window as well). This change in the anisotropy for the blue-MS
 sample can be related with the combined effect of rotation broadening affecting
 the $\mu_l$ distribution for fields close
to the Galactic minor axis (Zhao et al. 1996b) and disk contamination.

 All this evidence suggests that our method, even
   though does not completely isolate the bulge population
 kinematically, it is very useful to separate regions of the CMD
 where foreground disk contamination prevails. Furthermore, the fact 
 that the vertex angle decreases at higher latitudes gives us some 
 clues about the  extent of the dynamical influence of the bar feature in 
 the Galactic bulge. We are in the process of incorporating this information into
 a detailed dynamical model of the bulge/bar.

\section{Conclusions}
 We have described in this paper the procedures and results
 of a study which aims to identify a significant signature of the stellar
 bar in several windows with low foreground extinction in the Galactic bulge.
 Radial velocities have been derived from $\sim 110$ hours of VLT 
 VIMOS-IFU observations. 
 
 We have proven that our new method which combines the information from HST 
 photometry, proper motions, and IFU spectroscopy makes feasible the detection of
 the 3-D kinematics of bulge stars.
 The radial velocity procedure, based on a deconvolution in the spectral IFU cube 
 using HST positions and a IFU PSF have allowed us to obtain more than $\sim$3200 
 stellar radial velocities.
 
 Our large amount of data, combined with the proper motion information
 already presented in KR02 and K04, 
 have allowed the detection of a significant vertex deviation 
 in two of our three minor-axis fields. 
 Sagittarius-I and Baade's Window have a significant vertex deviation 
 while a weak signature is observed for the third minor axis field, NEAR NGC 6558. 
 The fact that the bar presents its strongest signature in the first two fields 
 decreasing in the lower latitude field NEAR NGC 6558, delivers valuable information
 about the extent of the bar feature and must be intrinsically related to the 
 detailed structure of the Galactic bulge.     
  
 This project is still in progress. We will soon add proper motions 
 for our three off-axis fields which will complete six fields with radial velocities
 and proper motions at different locations of the Galactic bulge. Four additional fields with radial velocities and 
 proper motion at negative longitudes have been planned. Thus, 
 our project will sample both ends of the bar obtaining robust constraints in the 
 characteristic bulge parameters. Moreover,
  a self consistent Schwarzschild model to disentangle the kinematic information 
 in these ten fields is under development. In the end, we expect to determine 
 a detailed picture of the stellar bar and its parameters using the 
 radial velocities and proper motions in our fields.

\begin{acknowledgements}
 We thank Dr. HongSheng Zhao for helpful discussions. Also, we would
 like to thank the referee of this work for useful comments and
 suggestions. MS 
 acknowledges support by \emph{Fondecyt} project No 3110188 and 
 Comit\'e Mixto ESO-Chile.
 Support to R. M. R. for proposals GO-8250, GO-9436, GO-9816 and GO-11655 was provided by NASA through grants from the Space Telescope Science Institute, which is operated by the Association of Universities for Research in Astronomy, Inc., under NASA contract NAS 5-26555.
\end{acknowledgements}

\end{document}